%
\let\useblackboard=\iftrue       
%
%
\newfam\black       
\input harvmac.tex       
%
\input epsf.tex       
\ifx\epsfbox\UnDeFiNeD\message{(NO epsf.tex, FIGURES WILL BE       
IGNORED)}       
\def\figin#1{\vskip2in}
\else\message{(FIGURES WILL BE INCLUDED)}\def\figin#1{#1}\fi       
\def\ifig#1#2#3{\xdef#1{fig.~\the\figno}       
\midinsert{\centerline{\figin{#3}}%
\smallskip\centerline{\vbox{\baselineskip12pt       
\advance\hsize by -1truein\noindent{\bf Fig.~\the\figno:} #2}}       
\bigskip}\endinsert\global\advance\figno by1}       
\noblackbox       
\def\Title#1#2{\rightline{#1}       
\ifx\answ\bigans\nopagenumbers\pageno0\vskip1in%
\baselineskip 15pt plus 1pt minus 1pt       
\else
\def\listrefs{\footatend\vskip       
1in\immediate\closeout\rfile\writestoppt       
\baselineskip=14pt\centerline{{\bf        
References}}\bigskip{\frenchspacing%
\parindent=20pt\escapechar=` \input        
refs.tmp\vfill\eject}\nonfrenchspacing}        
\pageno1\vskip.8in\fi \centerline{\titlefont #2}\vskip .5in}        
         
scaled\magstep3        
         
scaled\magstep3        
         
scaled\magstep3        
         
scaled\magstep3        
         
scaled\magstep3        
\ifx\answ\bigans\def\tcbreak#1{}\else\def\tcbreak#1{\cr&{#1}}\fi        

\useblackboard        
\message{If you do not have msbm (blackboard bold) fonts,}        
\message{change the option at the top of the tex file.}

\font\blackboard=msbm10 scaled \magstep1        
\font\blackboards=msbm7        
\font\blackboardss=msbm5        
\textfont\black=\blackboard        
\scriptfont\black=\blackboards        
\scriptscriptfont\black=\blackboardss        
        
\else        
        
\fi        

%
\def\yboxit#1#2{\vbox{\hrule height #1 \hbox{\vrule width #1        
\vbox{#2}\vrule width #1 }\hrule height #1 }}        
\def\fillbox#1{\hbox to #1{\vbox to #1{\vfil}\hfil}}        
\def\ybox{{\lower 1.3pt \yboxit{0.4pt}{\fillbox{8pt}}\hskip-0.2pt}}        

\def\comments#1{}

\def\half{{1\over 2}}

\def\CN{{\cal N}}        
\def\CP{{\cal P}}

\def\II{\relax{I\kern-.07em I}}        
\def\IIA{{\II}A}

\def\inbar{\,\vrule height1.5ex width.4pt depth0pt}        
\def\IZ{\relax\ifmmode\mathchoice        
{\hbox{\cmss Z\kern-.4em Z}}{\hbox{\cmss Z\kern-.4em Z}}        
{\lower.9pt\hbox{\cmsss Z\kern-.4em Z}}        
{\lower1.2pt\hbox{\cmsss Z\kern-.4em Z}}\else{\cmss Z\kern-.4em        
Z}\fi}        
\def\IB{\relax{\rm I\kern-.18em B}}        
\def\IC{{\relax\hbox{$\inbar\kern-.3em{\rm C}$}}}        
\def\ID{\relax{\rm I\kern-.18em D}}        
\def\IE{\relax{\rm I\kern-.18em E}}        
\def\IF{\relax{\rm I\kern-.18em F}}        
\def\IG{\relax\hbox{$\inbar\kern-.3em{\rm G}$}}        
\def\IGa{\relax\hbox{${\rm I}\kern-.18em\Gamma$}}        
\def\IH{\relax{\rm I\kern-.18em H}}        
\def\IK{\relax{\rm I\kern-.18em K}}        
\def\IP{\relax{\rm I\kern-.18em P}}        
\def\pp{{\relax{=\kern-.42em |\kern+.2em}}}        

\font\cmss=cmss10 \font\cmsss=cmss10 at 7pt        
\def\IR{\relax{\rm I\kern-.18em R}}

%


%
%
        
\def\NP{{\it Nucl. Phys.\ }}        
        
\def\PL{{\it Phys. Lett.\ }}        
\def\PR{{\it Phys. Rev.\ }}

\def\ATMP{{\it ATMP\ }}        

\Title{ \vbox{\baselineskip12pt\hbox{hep-th/9805158}        
\hbox{BROWN-HET-1125}\hbox{TUW-98-10}        
}}        
{\vbox{         
\centerline{Supersymmetric Gauge Theories from }        
\centerline{Branes and Orientifold         
Six-planes}}}        

\centerline{Karl Landsteiner$^{\natural}$,         
Esperanza Lopez$^{\natural}$        
 and David A. Lowe$^{\flat}$}     
\medskip        
\centerline{$^\natural$ Institut f\"ur theoretische Physik, TU-Wien}    
\centerline{Wiedner Hauptstra{\ss}e 8-10}        
\centerline{A-1040 Wien, Austria}        
\centerline{\tt landstei@tph45.tuwien.ac.at}        
\centerline{\tt elopez@tph16.tuwien.ac.at}        
\medskip        
\centerline{$^\flat$Department of Physics}        
\centerline{Brown University}        
\centerline{Providence, RI 02912, USA}        
\centerline{\tt lowe@het.brown.edu}        
\bigskip        
        
\centerline{\bf{Abstract}}        
        
We study brane configurations in the presence of orientifold six-planes.   
After deriving the curves for $\CN=2$ supersymmetric         
$SU(N_c)$ gauge theories with one         
flavor in the symmetric or antisymmetric representation and $N_f$      
fundamental        
flavors, we rotate the brane configuration, reducing the supersymmetry to         
$\CN=1$. For the case of an antisymmetric flavor and less than two  
fundamental        
flavors, nonperturbative effects lead to a brane configuration that        
is topologically a torus. Using the description of the orientifold         
six-planes as         
$D_n$ singularities we        
discuss the Higgs branches for $\CN=2$ brane configurations with  
$Sp$/$SO$ gauge groups and the related $\CN=1$ theories         
with tensor representations.         
\vfill        
\Date{\vbox{\hbox{\sl May, 1998}}}        
\lref\lalo{K. Landsteiner and E. Lopez, ``New Curves from Branes,''    
\NP {\bf B516} (1998) 273, hep-th/9708118.}        
\lref\lllii{K. Landsteiner, E. Lopez and D. A. Lowe, ``Duality of Chiral $\CN=1$         
Supersymmetric Gauge Theories via Branes,'' JHEP {\bf 02} (1998) 007, hep-th/9801002;    
I. Brunner, A. Hanany, A. Karch and D. L\"ust, ``Brane Dynamics and Chiral         
non-Chiral Transitions,'' hep-th/9801017;        
S. Elitzur, A. Giveon, D. Kutasov and D. Tsabar, ``Branes, Orientifolds and         
Chiral Gauge Theories,'' hep-th/9801020.}        
\lref\llli{K. Landsteiner, E. Lopez and D. A. Lowe, ``N=2 Supersymmetric Gauge         
Theories, Branes And Orientifolds,'' \NP {\bf B507} (1997) 197, hep-th/9705199.}         
\lref\witM{E. Witten, ``Solutions Of Four-Dimensional Field Theories Via M Theory,''     
\NP {\bf B500} (1997) 3, hep-th/9703166.}        
\lref\sp{A. Hanany, ``On the Quantum Moduli Space of N=2 Supersymmetric Gauge     
Theories,'' \NP {\bf B466} (1996) 85, hep-th/9509176;    
P. C. Argyres and A. D. Shapere, ``The Vacuum Structure of N=2     
SuperQCD with Classical Gauge     
Groups,'' \NP {\bf B461}(1996) 437, hep-th/9509175;    
E. D'Hoker, I. M. Krichever and D. H. Phong, ``The Effective Prepotential of     
N=2 Supersymmetric $SO(N_c)$ and $Sp(N_c)$ Gauge Theories,'' \NP {\bf B489} (1997)     
211, hep-th/9609145.}        
\lref\hov{K. Hori, H. Ooguri and C. Vafa, ``Nonabelian Conifold Transitions         
And N=4 Dualities In Three-Dimensions,'' \NP {\bf B504}    
 (1997) 147, hep-th/9705220.}        
\lref\dhoo{J. de Boer, K. Hori, H. Ooguri and Y. Oz, ``Branes and Dynamical         
Supersymmetry Breaking,'' hep-th/9801060.}        
\lref\csst{C. Csaki, M. Schmaltz, W. Skiba and J. Terning, ``Gauge Theories         
with Tensors from Branes and Orientifolds,'' hep-th/9801207.}        
\lref\radu{C. Ahn, K. Oh and R. Tatar, ``Comments On SO / Sp Gauge Theories         
From Brane Configurations With An O(6) Plane,'' hep-th/9803197.}  
\lref\givkut{A. Giveon and D. Kutasov, ``Brane Dynamics and Gauge Theory'',        
hep-th/9802067.}        
\lref\barbon{M. Berkooz, M.R. Douglas and R.G. Leigh, ``Branes Intersecting at      
Angles'', \NP {\bf B480} (1996) 265,     
hep-th/9606139;        
J.L.F. Barbon, ``Rotated Branes and $\CN=1$ Duality,''        
\PL {\bf B402} (1997), 59, hep-th/9703051.}        
\lref\jp{J. Park, ``M-theory realization of a N=1 supersymmetric chiral gauge         
theory in four dimensions,'' hep-th/9805029.}        
\lref\torused{E. Witten, ``Toroidal Compactification Without Vector Structure,''        
JHEP {\bf 02} (1998) 006, hep-th/9712028.}        
\lref\gipe{A. Giveon and O. Pelc, ``M Theory, Type IIA String and 4D N=1        
SUSY $SU(N_L) \times SU(N_R)$ Gauge Theory,'' \NP {\bf B512} (1998), 103,       
hep-th/9708168.}       
\lref\ssw{A. Sen, ``F-theory and Orientifolds,'' \NP {\bf B475} (1996),    
562, hep-th/9605150;     
N. Seiberg, ``IR Dynamics on Branes and Space-Time Geometry,''        
\PL {\bf B384} (1996), 81, hep-th/9606017;     
N. Seiberg and E. Witten, ``Gauge Dynamics And Compactification        
To Three Dimensions,'' hep-th/9607163.}        
\lref\uranga{M. Uranga, ''Towards Mass Deformed N=4 SO(n) and Sp(k) gauge        
configurations'', hep-th/9803054.}       
\lref\hoo{K. Hori, H. Ooguri and Y. Oz, ``Strong Coupling Dynamics of Four-     
Dimensional N=1 Gauge Theories from M Theory Fivebrane,'' \ATMP {\bf 1} (1998) 1,        
hep-th/9706082.}       
\lref\hawi{A. Hanany and E. Witten, ''Type IIB Superstrings, BPS Monopoles,      
And Three-Dimensional Gauge Dynamics'',     
\NP {\bf B492} (1997) 152, hep-th/9611230.}       
\lref\klwv{A. Klemm, W. Lerche, P. Mayr, C. Vafa and N. Warner, ''Self-Dual      
Strings and N=2 Supersymmetric Field Theory'', \NP {\bf B477} (1996) 746,     
hep-th/9604034.}     
\lref\nos{S. Nam, K. Oh and S.-J. Sin, ''Superpotentials of N=1      
Supersymmetric Gauge Theories from M-theory,'' \PL {\bf B416} (1998) 319,     
hep-th/9707247.}     
\lref\spanti{P. Cho and P. Kraus, ``Symplectic SUSY Gauge Theories with    
Antisymmetric Matter,'' \PR {\bf D54} (1996) 7640, hep-th/9607200;     
C. Csaki, W. Skiba and M. Schmaltz, ``Exact Results and Duality for SP(2N)      
SUSY Gauge Theories with an Antisymmetric Tensor,'' \NP {\bf B487}(1997) 128,     
hep-th/9607210.}      
\lref\evans{ N. Evans, C. V. Johnson and A. Shapere, ``Orientifolds,     
Branes and Duality of 4D Gauge Theories,'' \NP {\bf B505} (1997) 251,      
hep-th/9703210.}     
\lref\haza{A. Hanany and A. Zaffaroni, ''Chiral Symmetry from Type IIA Branes,'',    
\NP {\bf B509} (1998), 145, hep-th/9706047.}

\newsec{Introduction}        
        
The study of supersymmetric gauge theories in various dimensions       
realized by brane configurations has been a very active       
research area in recent times. Many new results concerning the       
non-perturbative behavior of gauge theories have been obtained.       
A recent review of the techniques involved and a summary of        
relevant references was given in \givkut.        
       
To study four-dimensional supersymmetric gauge theories one commonly       
uses the following building blocks. In type \IIA\ string theory one  
considers a collection of parallel fivebranes and fourbranes stretched     
between them. The world volume of the fourbranes is then bounded       
by the fivebranes and thus of finite extent in one direction (commonly     
labeled $x^6$). At sufficiently low energies       
the physics of the fourbranes is then described by a four-dimensional    
gauge theory. Such a configuration will leave eight supercharges       
unbroken giving rise to a four-dimensional $\CN=2$ gauge theory \witM.   
If one rotates the fivebranes such that the rotation lies in an       
$SU(2)$ subgroup of the $SO(4)$ rotation group transverse to the fivebranes,     
one breaks another half       
of the supersymmetries \barbon. In this way one can construct brane        
configurations       
corresponding to four-dimensional $\CN=1$ supersymmetric gauge theories.        
Sixbranes in between two fivebranes with an orientation that preserves     
the supersymmetries give rise to matter transforming in the fundamental    
representation of the gauge group. Upon going from type \IIA\ string    
theory to M-theory the fourbranes become themselves fivebranes that are    
wound around the eleventh dimension. The whole brane configuration becomes       
a single fivebrane with worldvolume $R^4\times \Sigma$ with $\Sigma$ being a        
Riemann surface \klwv. Information about the nonperturbative behavior of the       
gauge theory is encoded in this Riemann surface.        
       
An important ingredient in these brane constructions are       
orientifold planes. They can be introduced as four-orientifolds parallel   
to the fourbranes or as six-orientifolds parallel to the sixbranes.  
Both possibilities are compatible with supersymmetry and give rise       
to orthogonal or symplectic gauge groups. Six-orientifolds are of       
particular interest. It has been shown that by placing a six-orientifold   
on top of a fivebrane one obtains       
gauge theories with $SU(N_c)$ gauge group and matter transforming in       
the symmetric or antisymmetric representation. The fivebrane can       
also divide the six-orientifold in two and this gives rise to a       
chiral $\CN=1$ theory \lllii.        
       
This paper is devoted to the investigation of several aspects of        
brane configurations with six-orientifolds. In section two we use        
a configuration consisting of three parallel fivebranes and a        
six-orientifold on top of the central one. We also include a number  
of sixbranes. We obtain $\CN=2$ $SU(N_c)$ gauge theories with  one flavor        
of symmetric or antisymmetric matter and $N_f$ fundamentals. From        
the brane configurations we derive the curves parameterizing the       
Coulomb branch of these theories. We note that there is the interesting    
effect of non-perturbative mass generation for $N_f = N_c-3$ in the case      
with        
a symmetric flavor and for $N_f = N_c+1$ in the case with an antisymmetric       
flavor \foot{A similar effect was found in \gipe.}.       
       
In section three we rotate the outer fivebranes and break to $\CN=1$.    
The corresponding brane configuration in M-theory is parameterized by a    
$\CP^1$. However, it turns out that in the case with an antisymmetric    
flavor and one or zero fundamental flavors the brane configuration       
is not birational to a sphere. We go on to investigate these cases       
further in section four. There we argue that non-perturbative effects    
due to the orientifold generate an additional handle and that the        
brane configuration is actually a genus one curve. We show that the  
asymptotic behavior is consistent with the assumption of a genus one       
curve. In section five we briefly comment on the chiral brane configuration.       
Section six discusses the Higgs branches of $\CN=2$ brane configurations       
corresponding to orthogonal and symplectic gauge theories. We use there    
the description of six-orientifolds as $D_n$ singularities \ssw. It was    
suggested       
in \torused\ that the six-orientifold giving rise to orthogonal gauge groups       
can be described by an $D_{N_f+4}$ singularity. It is important that in    
this case the singularity can only be resolved down to $D_4$. We compute the       
dimension of the Higgs branch using this description and show that it    
indeed coincides with field theory. In section seven we discuss the        
Higgs branches of $\CN=1$ $SO$/$Sp$ gauge theories with tensor      
representations.       
Again using the description of the six-orientifold as a $D_n$ singularity we      
can       
compute the dimensions of the various Higgs branches.       
       
After this work was completed we learned of independent work \jp\ that     
partially overlaps with some results in sections three, five and six.

\newsec{Curves for $\CN=2$ $SU(N_c)$ with a tensor flavor and         
$N_f$ fundamentals}        
\ifig\figi{An $\CN=2$ brane configuration with three fivebranes. On top of      
the        
middle fivebrane there is a six-orientifold. Such configurations give rise      
to        
$SU(N_c)$ gauge theories with matter hypermultiplets either in the symmetric      
or        
antisymmetric representation depending on the sixbrane charge of the  
orientifold.        
In addition there are some sixbranes        
giving rise to hypermultiplets in the fundamental representation.}  
{
\epsfxsize=3.5truein\epsfysize=3truein        
\epsfbox{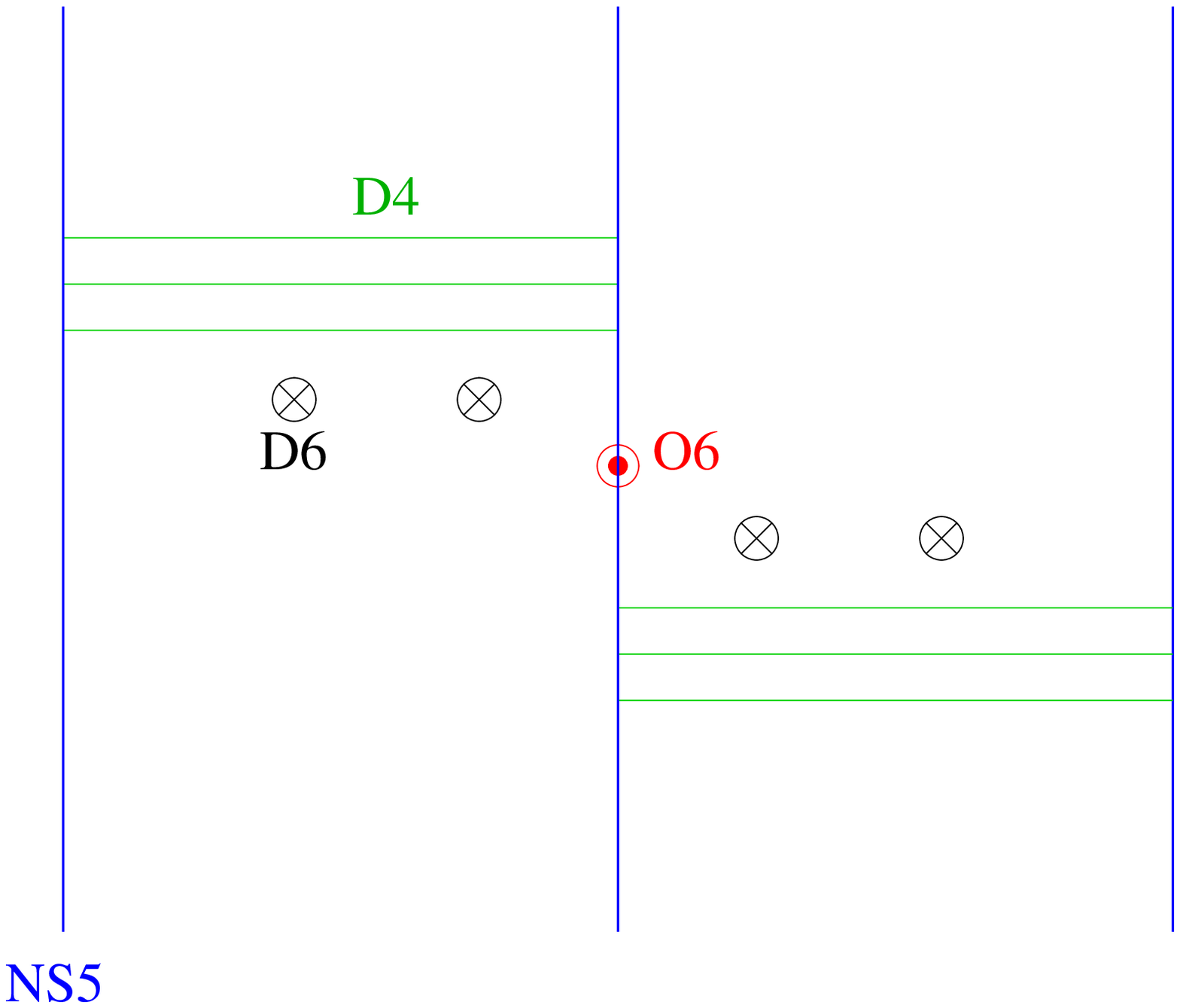}}        
\subsec{Symmetric flavor}        
        
In this section, the Seiberg-Witten curve for an $\CN=2$         
$SU(N_c)$ gauge theory with a flavor of symmetric and $N_f$ fundamental    
flavors is        
constructed.        
Following \witM, we will derive         
it by lifting a certain type \IIA\ brane configuration to M-theory.        
The basic brane configuration is that considered in \lalo: three NS        
fivebranes ($012345$) and $N_c$ fourbranes suspended between them ($01236$),        
in the presence of an orientifold sixplane of $+4$ Ramond charge ($0123789$).        
We have indicated in brackets the type \IIA\ directions in which each object        
extends. There will        
be in addition $2N_f$ sixbranes parallel to the orientifold (see \figi).   
In order to describe the lifting to M-theory of this configuration, it is        
common to introduce complex coordinates $v=x_4+ i x_4$ and $s=(x_6 +   
i x_{10})/R$ where $x_{10}$ denotes the eleventh dimension of M-theory and        
$R$ is its radius. A orientifold sixplane of positive Ramond charge and    
$2N_f$ sixbranes can be described in M-theory by \lalo       
\eqn\xyspace{        
xy= (-1)^{N_f} v^4 \prod_{k=1}^{N_f} (v^2 - e_k^2),        
}        
where $v^4$ comes from the orientifold six-plane, and $x,y$ are complex    
coordinates such that for small or fixed $x$, $y\sim e^{-s}$, and for small        
or fixed $y$, $x\sim e^s$. The parameters $e_k$ correspond to the position       
of the sixbranes in the $v$-plane \witM. The orientifold action in these   
coordinates is $(x,y,v) \rightarrow (y,x,-v)$. Finally, it is convenient   
to define         
\eqn\quarkpol{        
\eqalign{        
j_1(v) & =   \prod_{k=1}^{N_f} (v-e_k), \cr        
j_2(v) & = (-1)^{N_f} \prod_{k=1}^{N_f} (v+e_k) . \cr}}        
        
The M-theory curve describing a brane configuration with several        
fivebranes in a background space of the form \xyspace\ was given        
in \witM. We just have to impose in addition invariance under the        
orientifold projection. We obtain the following curve for the        
configuration with three fivebranes        
\eqn\symF{        
y^3 + y^2 p(v) + y v^2 j_1(v) q(v) +  v^6 j_1^2 (v) j_2(v)=0,        
}        
where $q(v)=p(-v)$ and $p(v)=\prod_{i=1}^{N_c} (v-a_i)=v^{N_c} +   
{\half N_c m  } v^{N_c-1} + u_2 v^{N_c-2} + ...$. The parameters $a_i$     
represent the positions of the fourbranes on the $v$-plane.        
Combinations of them give $m$, the mass of the symmetric flavor,        
and $u_k$, $k=2,\cdots,N_c$, the $SU(N_c)$ casimirs. The sixbranes       
induce $N_f$ fundamental hypermultiplets for the $SU(N_c)$ gauge       
theory. Notice that the masses of these hypermultiplets are given       
by (see \figi)        
\eqn\masses{       
m_k={m \over 2} -e_k.       
}       
        
The $SU(N_c)$ theory with a symmetric and $N_f$ fundamental flavors  
has baryonic operators $B_n= X^n Q^{N_c -n} Q^{N_c -n}$,        
${\tilde B}_n= {\tilde X}^n {\tilde Q}^{N_c -n} {\tilde Q}^{N_c -n}$,    
where $X,{\tilde X}$ represent the fields in the symmetric         
representation and $Q,{\tilde Q}$ the quarks.        
When $B_n$ or ${\tilde B}_n$ get an expectation value the initial        
$SU(N_c)$ theory breaks to $SO(n)$ with $N_f-(N_c-n)$ flavors.        
In the brane language, the baryonic branches will correspond to         
factorizing the central fivebrane. The curve \symF\ factorizes into        
\eqn\symfacE{        
(y+ v^2 j_1)(y^2 + y (p-v^2 j_1) +   v^4 j_1 j_2)=0~,        
}        
when        
\eqn\symfaccondE{        
p(v)- v^2 j_1(v) = q(v) - v^2 j_2(v)~,        
}        
which implies that $p-v^2 j_1$ only contains         
even powers of $v$. The second factor in \symfacE\ corresponds         
to the curve of an $SO(N_c)$ $\CN=2$ gauge theory with         
$N_f$ flavors \sp.        
This is the expected breaking associated with expectation values        
for $B_{N_c}$ and ${\tilde B}_{N_c}$.         
        
The first term in \symfacE\ represents the factorized central fivebrane.   
From field theory we get that the quarternionic dimension of this
baryonic branch is one. In the brane language three of the four
real parameters can be related to the $x_7,x_8,x_9$ position of the
fivebrane. The field theory has a global $U(1)_X$ symmetry acting
only on the field in the two-index tensor representation.
The fourth parameter can be understood as the Goldstone mode for 
this $U(1)_X$ symmetry. However, this $U(1)_X$ is not realized
geometrically in the brane configuration. Therefore we can not see the 
fourth real parameter as a geometric quantity. It should be related 
to the integration of the chiral antisymmetric two-tensor field along 
the world-volume, when suitably regularized.

        
For $N_c$ odd the curve \symF\ factorizes as        
\eqn\symfacO{        
(y- v^2 j_1)(y^2 + y (p+v^2 j_1) -  v^4 j_1 j_2)=0~,        
}        
when        
\eqn\symfaccondO{        
p(v)+ v^2 j_1(v) = -(q(v) +  v^2 j_2(v))~,        
}        
which in this case implies $p+v^2 j_1 = v B(v^2)$. The        
second factor in \symfacO\ describes a brane configuration        
with two fivebranes and an odd number of fourbranes         
in the space \xyspace. By redefining        
$y \rightarrow v y$ we obtain the Seiberg-Witten curve        
associated with an odd orthogonal theory with $N_f$ flavors in its     
standard form \sp \foot{The redefinition $y \rightarrow vy$ does not       
have meaning in the M-theory context. It would be       
equivalent to resolve part of the $D_4$ singularity associated        
with the orientifold. This is however not allowed \torused, \lalo.}.    
        
Notice that the conditions for factorization \symfacE\        
and \symfacO\ are not simply $q=p$ and $q=-p$ for $N_c$ even        
and odd respectively, as one would have naively expected.        
In particular for $N_f=N_c-3$ \symfacE\ and \symfacO\       
imply that       
\eqn\npmass{       
m \sim {\Lambda_{\CN=2}},       
}       
where we have restored the dependence on the dynamical scale       
of the theory, which we had set to $1$ in the above.       
An analogous shift in the mass of a bifundamental flavor was        
encountered when analyzing the Seiberg-Witten       
curves for an $SU(N_c) \times SU(N_c)$ gauge theory \gipe.       
As in \gipe\ we interpret this as the non-perturbative       
generation of a mass for the symmetric flavor, which has to        
be canceled for the factorization to occur. It also means that the location 
of the root of the        
baryonic branch where the middle fivebrane detaches suffers a        
non-perturbative correction. A somewhat analogous shift of a Higgs       
branch root will appear for the $N_f/2$-th branch of $\CN=2$ $Sp$  
theories in section six. It was also noted already in \dhoo. Notice that       
\masses\ suggests that also the quarks receive non-perturbative       
corrections to their mass.

The other baryonic branches appear when in addition to the       
factorization \symfacE\ or \symfacO, we also factorize $p(v) \pm        
v^2 j_1(v)$ as $v^{N_c-n} \tilde p(v)$. This amounts to putting the        
$SO(N_c)$ gauge theory at the origin of its $(N_c-n)$-Higgs branch.        
We may then blow up the $D_{N_f+4}$ singularity as described in section 6,        
taking us onto the Higgs branch of the $SO(N_c)$ theory. In this way we    
obtain exactly the number of moduli necessary to describe the $B_n$ and    
$\tilde B_n$ baryonic branches of the original $SU(N_c)$ theory, since     
their quaternionic dimension equals the dimension of the $(N_c-n)$-Higgs   
branch of an $SO$ theory with $N_f$ flavors, plus one. The additional      
one corresponds to the factorized central fivebrane as before.        
        
\subsec{Antisymmetric flavor}        
        
We consider now the same brane configuration as in the previous         
section, but in the background of an orientifold sixplane of $-4$        
Ramond charge. As explained in \lalo\ this configuration induces an        
$\CN=2$ $SU(N_c)$ gauge theory with an antisymmetric flavor on        
the world-volume of the fourbranes. We include again the presence of       
$2N_f$ sixbranes, which will provide $N_f$ fundamental flavors for         
the $SU(N_c)$ theory. An orientifold sixplane of negative charge and       
$2N_f$ coincident sixbranes are described in M-theory by         
compactification in a complex two-dimensional space with a $D_{N_f}$       
singularity \ssw. The sixbranes can be taken in pairs away        
from the orientifold. As a complex manifold, this is represented by a      
deformed $D_{N_f}$ surface \hov        
\eqn\dnspace{        
a^2 + b^2 z = {4  \over z} \left( \prod_{k=1}^{N_f}         
(z+e_k^2) - \prod_{k=1}^{N_f} e_k^2 \right) -         
4 b \prod_{k=1}^{N_f} e_k ~.        
}        
The parameters $e_k$ describe the position of the sixbranes in the  
$z$ direction.         
        
As in \lalo, instead of \dnspace\ we will use a space that is         
birationally equivalent to it and which only provides a complete         
description of \dnspace\ far from the orientifold         
\eqn\xyvdn{        
xy= (-1)^{N_f} v^{-4} \prod_{k=1}^{N_f} (v^2 - e_k^2)~.        
}        
In this space we impose invariance under the orientifold         
projection $(y,x,v) \rightarrow (x,y,-v)$. The different spaces are related by        
$a = v(y-x)$, $b = x + y + 2 v^{-2} \prod_{k=1}^{N_f} e_k$,        
$z = - v^2$, which corresponds to choosing $\IZ_2$ invariant coordinates        
in \xyvdn.        
        
In the auxiliary space \xyvdn, the most general Riemann surface         
associated with a configuration of three fivebranes is        
\eqn\antisymF{        
y^3 + y^2 (p + B v^{-1} \! + 3 A v^{-2}) + y  v^{-2} j_1          
(q - B v^{-1} \! + 3 A v^{-2}) +  v^{-6} j_1^2  j_2=0~,        
}        
where again $q(v)=p(-v)$ and $p(v)=\prod_{i=1}^{N_c} (v-a_i)$, with $N_c$        
the number of         
fourbranes suspended between the fivebranes. The polynomials         
$j_1(v),j_2(v)$ are defined as in \quarkpol. The terms       
$B v^{-1}$, $3 A v^{-2}$ are not related with fourbranes positions.        
They are allowed by the presence of negative powers of $v$        
in \xyvdn. However the fact that they can not be fixed with the        
information contained in \xyvdn, is a sign that this space does         
not provide a good description of the orientifold background close to      
the origin. We use the space \xyvdn\ as a tool for obtaining a very    
restricted ansatz for the desired curve \lalo. Once we have the ansatz,    
the extra coefficients are determined by imposing that the curve can be    
written as a polynomial in the standard $D_{N_f}$ surface \dnspace, where       
no negative powers of $v$ appear.        
The result is        
\eqn\np{        
B= \Lambda_{\CN=2}^{N_c+2-N_f} \sum_{k=1}^{N_f} \prod_{l \neq k} (-e_l)  ~,        
\;\;\;\;\,\; A= \Lambda_{\CN=2}^{N_c+2-N_f} \prod_{k=1}^{N_f} (-e_k)~,       
}        
where we have restored the dependence on the dynamical scale       
$\Lambda_{\CN=2}$ of the $SU(N_c)$ theory.\foot{We thank I. Ennes,
S. Naculich, H. Rhedin and H. Schnitzer for pointing out a sign error
in an earlier draft.}
Both coefficients are proportional to the dynamical scale, indicating    
their origin in the strong coupling dynamics close to the orientifold.     
The parameters $e_k$ are again related to quark masses by \masses.       
        
As a check of the proposed curves we will analyze deformations of        
the previous curve associated with baryonic branches of the $\CN=2$ $SU(N_c)$        
theory with an antisymmetric flavor and $N_f$ quark flavors. This theory contains baryons         
of the form $B_n = X^n Q^{N_c-2n}$ and ${\tilde B}_n = {\tilde X}^n {\tilde Q}^{N_c-2n}$.        
When $N_c$ is even the baryons $B_{N_c}= X^{N_c/2}$,        
${\tilde B}_{N_c}= {\tilde X}^{N_c/2}$, break the initial        
theory down to an $Sp(N_c)$ theory with $N_f$ hypers in the         
fundamental representation. In agreement with this, \antisymF\        
factorizes as        
\eqn\antisymfacE{        
(y+ v^{-2} j_1)(y^2 + y (p-          
{\tilde j}_1 + 2 A v^{-2}) + v^{-4} j_1 j_2)=0~,        
}        
when        
\eqn\antisymfaccondE{        
p(v) - {\tilde j}_1 (v)= q(v) - {\tilde j}_2(v)~.        
}        
We have defined ${\tilde j}_2(v) ={\tilde j}_1(-v)$ and         
${\tilde j}_1(v) = v^{-2} (j_1(v) - B v - A)$. The factorization  
condition insures that only even powers of $v$ appear in the second        
factor of \antisymfacE, and we obtain the curve for an $Sp(N_c)$ $\CN=2$         
theory with $N_f$ flavors \sp\ as expected. This factorization     
property can alternatively be used to fix the coefficients $A$ and $B$.    
        
When $N_c$ is odd the highest baryon operators are $B_{N_c-1}=         
X^{(N_c-1)/2} Q$, ${\tilde B}_{N_c-1}= {\tilde X}^{(N_c-1)/2} {\tilde Q}$,        
which break the $SU(N_c)$ theory to $Sp(N_c-1)$ with $N_f-1$ flavors.      
According to this, for $e_k \neq 0$ and $N_c$ odd,         
the curve \antisymF\ does not factorize for any value of the casimirs.     
However when at least one $e_k=0$, we have $A=0$ and the         
curve factorizes as        
\eqn\antisymfacO{        
(y-v^{-2} j_1)(y^2 + y (p+        
{\tilde j}_1 + 2B v^{-1}) - v^{-4} j_1 j_2)=0~,        
}        
provided that        
\eqn\antisymfaccondO{        
p(v) +{\tilde j}_1 (v)= -(q(v) + {\tilde j}_2(v))~.        
}        
The conditions \antisymfaccondE, \antisymfaccondO\ for $N_f<N_c+1$     
imply that the antisymmetric flavor must be massless for the factorization      
to occur. For $N_f=N_c+1$ we have again a non-perturbative shift in the mass      
of the antisymmetric (see \npmass).  Thus when \antisymfaccondO\       
is fulfilled, $e_k=0$ implies a massless fundamental if $N_f<N_c+1$ and        
$m_k\sim\Lambda_{\CN=2}$ when $N_f=N_c+1$.          
      
\ifig\figii{Baryonic branch in the case $N_c$ odd.}        
{
\epsfxsize=3.5truein\epsfysize=3truein        
\epsfbox{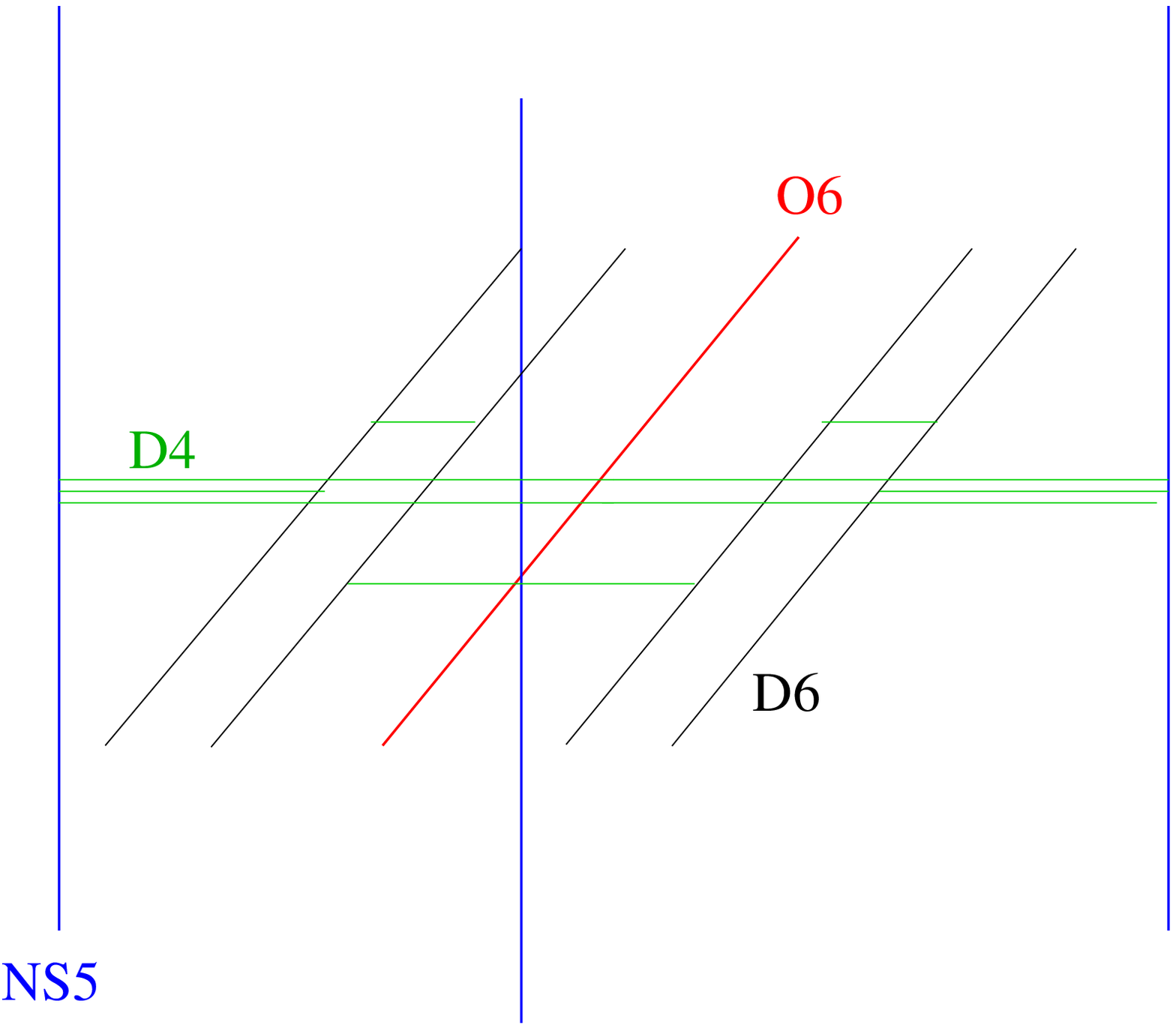}}     
      
From condition \antisymfacO\ we get $p(v) +  {\tilde j}_1 (v)= v b(v^2)$.        
Since one of the $e_k$ must be zero we have $j_1(v)=v        
j'_1(v)$, where $j'_1$ is the polynomial associated with a background space        
of $N_f-1$ sixbranes. Redefining then       
$y \rightarrow v y$, the second factor in \antisymfacO\ reproduces the     
Seiberg-Witten curve for $Sp(N_c-1)$ theory with $N_f-1$        
flavors.        
In this case we interpret the first factor in \antisymfacO\ as        
corresponding to the factorized central fivebrane with a fourbrane  
attached to it. One of the fourbranes has to remain attached to the      
middle fivebrane otherwise we would have an odd number of fourbranes  
intersecting an orientifold that projects onto symplectic groups, which    
is not consistent.     
The quaternionic dimension of the $B_{N_c-1}$, ${\tilde B}_{N_c-1}$      
baryonic branches is $N_f$. This is indeed correctly reproduced by the brane     
configuration. One modulus corresponds to the position of the central      
fivebrane together with the attached fourbrane and $N_f-1$ to the possible      
$\IZ_2$-symmetric breakings of this attached fourbrane in the $2N_f$  
sixbranes (see \figii).

\newsec{Curves for $\CN=1$ $SU(N_c)$  with a tensor flavor and $N_f$ fundamentals}        
        
Now we want to obtain the curves for the theories with $\CN =1$ supersymmetry. A way to achieve       
this is to introduce a mass for the chiral multiplet in the adjoint        
representation of the $\CN=2$ theory. As is well-known, in the the brane       
configuration this corresponds to rotating fivebranes from their original       
orientation along the $4,5$ directions towards the $8,9$ plane.       
We introduce a new complex variable $w=x^8 + i x^9$.       
In our case we have three fivebranes, of which the outer two can be rotated       
in a $\IZ_2$ symmetric manner. Thus the left fivebrane will be described   
asymptotically for large values of $v$ by ${v/ w} = \mu$ and the right one       
by ${v/w}=-\mu$. Furthermore we are interested in the case when the    
curve can be parameterized rationally by a $\CP^1$ whose coordinate we denote       
by $\lambda$. We will furthermore restrict ourselves to the case when all       
additional sixbranes lie at $v=0$ for the moment.     
\ifig\figiii{Projection onto the $(v,w)$-plane of the rotated configuration.}        
{
\epsfxsize=2.5truein\epsfysize=2.5truein        
\epsfbox{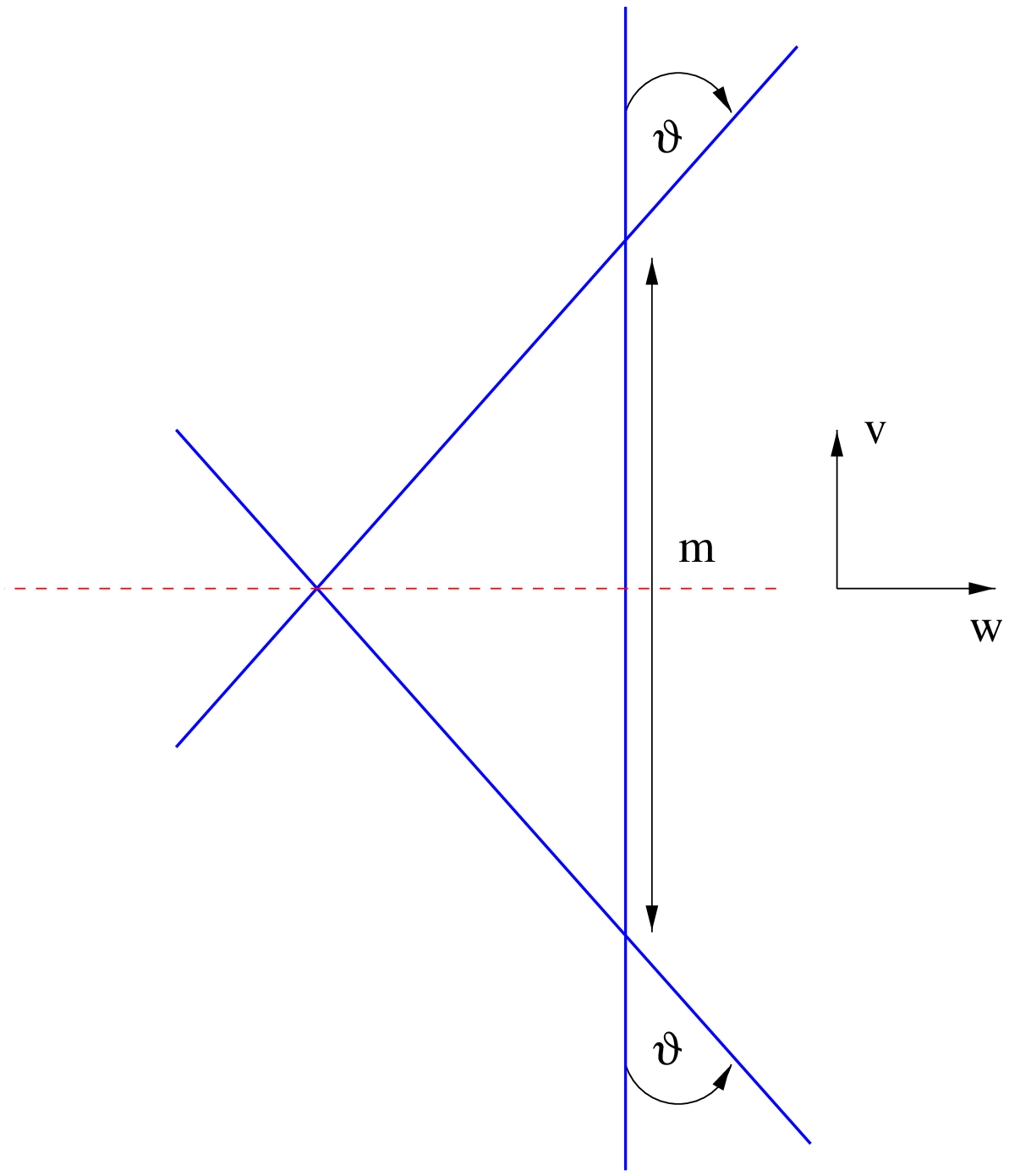}}      
The tree level superpotential associated with the rotated brane configuration     
can be obtained from the $\CN=2$ superpotential by integrating     
out the massive adjoint multiplet. The result is     
\eqn\w{     
W_{tree}= -{1 \over 2 \mu} \left( (X{\tilde X})^2+     
Q {\tilde X} X {\tilde Q} + (Q {\tilde Q})^2 \right)     
+ m X {\tilde X} + {m \over 2} Q {\tilde Q},     
}     
with $m$ as depicted in \figiii.          
       
The parameter $\mu$ is the only one carrying $U(1)_{89}$-charge associated with        
rotations in the $89$ plane. The $\CN=1$       
curve can be thought of as a deformation of the $\CN=2$ curve. However, because       
of the charge $\mu$ is carrying, it can not appear in the projection of the       
$\CN=1$ curve onto the the $(y,v)$-plane \hoo. It follows that this projection has       
the same form as the $\CN=2$ curve. Consider now the curves for the theories       
with a symmetric flavor and $N_f$ fundamentals and the theories with       
an antisymmetric flavor and $N_f+2$ flavors. From the expressions we derived       
in the previous section one sees that, by appropriate rescalings of $y$,   
these curves can be brought into the form        
\eqn\Nonecurve{v^M y^3 + y^2 p(v) + y p(-v) + (-1)^{M} v^M =0\,.}  
This allows us to treat the cases for the antisymmetric and     
the symmetric tensor at once. The orientifold is       
now assumed to act as $(y,v) \rightarrow (1/y,-v)$. We define $M=N_f+2$ in        
case of the symmetric and $M=N_f-2$ in case of the antisymmetric  
representation. In these coordinates one sees $M$ semi-infinite fourbranes       
to the left and to the right of the brane configuration located at $v=0$.       
It is important to note that in the case of the symmetric flavor two of    
these fourbranes represent the effects of the orientifold rather than      
matter in the fundamental representation \uranga. The cases $N_f=0,1$    
and an antisymmetric flavor are special. We postpone their study until     
further in this section.

Now we want to investigate the rotated brane configurations \nos.       
In terms of $\lambda$ we can assume       
\eqn\parametrization{\eqalign{ v =& {b \over \lambda -1} + {\tilde b \over        
\lambda} + { b \over \lambda +1}\,, \cr       
w =& {b\, \mu \over \lambda -1} - { b\, \mu \over \lambda +1} + 2 b\, \mu\,, \cr       
y =& A \left({\lambda -1 \over \lambda+1}\right)^{N_c-M}~.}}       
The form of $v$ follows since we have three fivebranes and correspondingly we       
demand three poles of order one for $v$ as a function of $\lambda$. Two of the       
fivebranes are interchanged by the action of the orientifold. On the sphere       
this orientifold acts by $\lambda \rightarrow -\lambda$. Since the middle     
fivebrane does not have a mirror image, it must approach asymptotically    
one of the two fixed points       
$v=\infty,~y=\pm 1$. The middle fivebrane must therefore be represented by a        
pole at $\lambda=0$ (equivalently we could choose $\lambda=\infty$). Without        
loss of generality,       
the positions of the poles for the outer fivebranes are set to       
$\lambda = \pm 1$. The coefficients of these poles are forced to be equal by       
the orientifold $\IZ_2$.        
       
The form of $w$ is given by similar considerations and       
by the asymptotic conditions ${w } = \pm \mu v$ at $\lambda = \pm 1$.       
We also included a shift in $w$ such that on the middle fivebrane $w$      
approaches zero for large values of $v$. Using the form \Nonecurve\ it  
follows that $y$ has a zero of order $N_c-M$ at $\lambda=1$, a pole of oder       
$N_c-M$ at $\lambda=-1$ and goes to a constant for $\lambda \rightarrow 0$.       
       
The constants $A,b,\tilde b$ can be fixed by inserting \parametrization\   
into \Nonecurve\ and examining the asymptotics at $\lambda=1$ and $\lambda=0$       
(the asymptotic expansion around $\lambda = -1$ is equivalent to the one at       
$\lambda=1$).        
Around $\lambda=1$ we have to leading and first subleading  order  
\eqn\asymone{\eqalign{ {1\over {(\lambda-1)}^M} &\Bigl( (-1)^{N_c} {A b^{N_c} \over 2^{{N_c}-M}} +        
(-1)^{N_c} b^M \Bigr)  +   \cr & +    
{1\over {(\lambda-1)}^{M-1}} \Bigl( (-1)^{N_c} {A b^{{N_c}-1}\over  
2^{{N_c}-M+1}} (b M + 2 \tilde{b} {N_c} - {N_c} m) +  \cr &+ (-1)^M M b^{M-1}({b\over2}       
+\tilde b ) \Bigr) + A^2 b^{N_c} {(\lambda-1)}^{{N_c}-2M} + \cdots =0\,.}}       
For a theory with asymptotic freedom we must have ${N_c} > M$, which has   
been        
assumed in \asymone.       
The last        
term contributes to the subleading order for ${N_c}=M+1$.      
     
Around $\lambda=0$ we       
have       
\eqn\asymzero{\eqalign{ {1\over \lambda^{N_c}} \Bigl((-1)^{2{N_c}-2M} A^2 \tilde{b}^{N_c} + (-1)^{2{N_c}-M} A        
\tilde{b}^{N_c}\Bigr) + {1\over\lambda^{{N_c}-1}} \Bigl( 4(-1)^{2{N_c}-2M} A^2 \tilde{b}^{N_c}(M-{N_c}) +\cr       
+ {{N_c} m\over 2} (-1)^{2{N_c}-2M} A^2 \tilde{b}^{{N_c}-1} +       
2(-1)^{2{N_c}-M} A \tilde{b}^{N_c}(M-{N_c}) -        
{{N_c} m\over 2} (-1)^{2{N_c}-M} A \tilde{b}^{{N_c}-1}\Bigr) + \cr       
+ {1\over \lambda^M} \Bigl((-1)^{3{N_c}-3M} A^3 \tilde{b}^M + (-1)^M \tilde{b}^M\Bigr) +      
\cdots =0~.}}        
Again, the last term contributes to the subleading order only for        
${N_c}=M+1$. The terms at lower orders do not give new conditions for the       
constants in \parametrization\ but determine the casimirs $u_i$. These equations are solved by       
\eqn\solutions{\eqalign{ A =& (-1)^{M+1} \Lambda_{\CN=2}^{N_c-M} \,,\cr       
b^{{N_c}-M} =& (-1)^{N_c} (2\Lambda_{\CN=2})^{{N_c}-M}  \,,\cr       
\tilde{b} =&{ {N_c} m\over 2({N_c}-M)} \,\qquad\;\hbox{if }\;{N_c} > M+1 \,,\cr       
\tilde{b} =& (-1)^M\Lambda_{\CN=2} + {{N_c} m\over 2} \,\quad\hbox{if }\;{N_c} = M+1 \,.       
}}       
Again we restored the dependence on the scale $\Lambda_{\CN=2}$.    
It is interesting to note that the coefficient for the middle fivebrane    
is proportional to the mass $m$. This also means that for zero mass  
the pole of $v$ corresponding to the middle fivebrane is absent, which implies the curve       
factorizes. This is in line with field theory expectations -- the       
baryonic branches will open up when the mass of the tensor vanishes.    
Therefore       
for zero mass       
with the central fivebrane factorized the curves are the same as the corresponding       
curves for $SO$ or $Sp$ theories with tensor matter \refs{\csst,\radu}.    
       
For ${N_c}=M+1$ however, the coefficient       
for the middle fivebrane vanishes when $m = (-1)^{M+1} {2\Lambda_{\CN=2}/    
{N_c}}$. This is of       
course the same non-perturbative shift in the mass as we noted       
in the previous section.

Let us analyze the behavior of the solutions at $v=0$. In terms of $\lambda$       
this corresponds to the three points $\lambda=\infty$ and $\lambda = \pm       
\sqrt{\tilde b / (2b+\tilde b)}$. For generic $\tilde b$ we therefore    
get three different values $y=(A,y_+,y_-)$. This means that the       
projection of the curve to the $(y,v)$ plane at $v=0$ is        
given by $(y-A)(y-y_+)(y-y_-)=0$. This is possible only if the polynomial       
$p(v)$ factorizes as $p(v) = v^M \tilde p(v)$. In the coordinates  
we are using we have $M$ semi-infinite fourbranes to the left and to the   
right of the brane configuration. The particular form of $p(v)$ means that       
we first have to bring $M$ fourbranes to $v=0$ and reconnect them with   
the semi-infinite ones. Only then we can perform the rotation to $\CN=1$.        

We will study now the cases with $N_f=0,1$ and an antisymmetric flavor.        
Instead of \Nonecurve, the $(y,v)$ projection of the curve for $N_f=1$ is        
given by       
\eqn\antisymNfone{y^3 + y^2 (v p(v) +1) + y (v p(-v) -1) -1=0\,,}  
and for $N_f=0$ by       
\eqn\antisym{y^3 + y^2 (v^2 p(v) +3) + y (v^2 p(-v) +3) +1=0\,.}       
We consider first $N_f=1$. The asymptotic conditions        
can be taken over from \asymone\ and \asymzero. This would lead to a    
similar curve        
as found above.       
However, now we have to take into account that        
the curve takes a definite form at $v=0$ and we can not adjust some    
order parameter to solve the equation. Indeed for $v=0$ we need to       
satisfy $(y-1)(y+1)^2 =0$. The zeroes of $v$ are given as above. For     
$\lambda = \infty$, $y=-1$ since from the asymptotic conditions        
$A=-1$. For generic $\tilde b$ the other two solutions of $v=0$ will   
not give $y=\pm 1$.       
Moreover, we already have one solution at $y=-1$. We would need to have       
two more solutions $\lambda_\pm$ with $y=\pm1$. The        
orientifold action exchanges $\lambda_+$ with $\lambda_-$ but leaves       
the $y$-values $\pm 1$ fixed. Thus if $y(\lambda_+) = 1$ then also    
$y(\lambda_-)=1$. This argument shows that we can not achieve the required       
structure of the solutions for $y$ through the parameterization by a sphere.       
We will argue in the next section that the curves in this case are  
topologically a torus rather than a sphere. When the mass of the        
antisymmetric vanishes the middle fivebrane can again factorize       
as described in       
the previous section. Up to the detached middle fivebrane the curve can    
be rotated to $\CN=1$ in the same way as the curves for $Sp(N_c)$ when   
$N_c$ is even or for $Sp(N_c-1)$ when $N_c$ is odd.       
       
Consider now the case       
with one antisymmetric flavor and       
no fundamentals at all. Again we have the asymptotic conditions       
\asymone\ and \asymzero, implying $A=-1$, $b^{{N_c}+2} = (-2)^{{N_c}+2}$ and       
$\tilde b = {m / 2}$. Also here we have to investigate carefully the     
behavior at $v=0$. This time the $(y,v)$ projection give rise to the     
triple point $(y+1)^3=0$. Now all the zeroes of $v$ in $\lambda$ have    
to give $y=-1$. Thus we have the equation       
\eqn\specialzeroesequ{       
\left( {\lambda - 1 \over \lambda +1} \right)^{{N_c}+2} = 1\,,}       
with the solutions       
\eqn\specialzeroes{\lambda = i \cot \left( {n \pi \over  N_c+2}     
\right)\,,}       
where $n=0,\cdots,{N_c}+1$. $n=0$ corresponds to $\lambda=\infty$, the other       
two solutions $\lambda_\pm$ come as the pair $n,{N_c}+2-n$ for fixed $n$.        
Note that this also implies specific values for the mass $m$ of the     
antisymmetric. In the next section we will argue that for generic $m$  
the curve is again topologically equivalent to a torus.     
For ${N_c}$ even and       
$n={N_c}/2+1$ we get $\lambda_\pm =0$. This implies $\tilde b = m =0$.       
This is consistent with the fact that only for even ${N_c}$ the middle  
fivebrane        
can factorize. For this case the other component of     
the curve will be equivalent to that of the $Sp$ gauge theory that appears      
on the baryonic branch.        
    
Finally, let us consider the form of the curve with an (anti)symmetric   
flavor and with nontrivial masses for the fundamental flavors. We switch   
back to a description where the ambient space is given by \xyvdn. For   
simplicity, we will set the mass of the tensor to zero,   
which corresponds to working at a point where the $\CN=2$ curve       
factorizes \foot{For $N_c = M+1$ and $M$ defined as before, we should fix 
the mass of   
the (anti)symmetric flavor to the quantum corrected value $m=(-1)^{N_f+1}   
{2\Lambda_{\CN=2} \over N_c}.$}.   
We will also take ${N_c}$ to be even, thus the resulting curves   
will also describe $SO(2n_c)$ and $Sp(2n_c)$ gauge theories where $N_c=2n_c$.

With an antisymmetric flavor the curve takes the form       
\eqn\vthing{       
\eqalign{       
v &= {b\over {\lambda-1}} +  {b\over {\lambda+1}} \cr       
w &= {\mu b\over {\lambda-1}} -  {\mu b\over {\lambda+1}} \cr       
y &= A v^{-2} \prod_{i=1}^{N_f} {\lambda^2 + 1 - \lambda c_i \over  
\lambda^2 -1} \left( {\lambda-1 \over \lambda+1} \right)^{2n_c+2-N_f} ~,\cr       
}}       
where we have defined       
\eqn\wehdef{       
\eqalign{       
c_i^2 &= {{4(b^2+m_i^2)}\over m_i^2} \cr       
b^{4n_c+4} &= (2\Lambda_{\CN=2})^{4n_c+4-2N_f}      
\prod_{i=1}^{N_f} m_i^2(1+c_i/2)^2\cr       
A^2 &= \Lambda_{\CN=2}^{4n_c+4-2N_f}\prod_{i=1}^{N_f} m_i^2 ~,\cr}   
}  
with $m_i$ the masses of the fundamental flavors.       
The form of \vthing\ is determined by asking that $x+y=p(v^2)$.       
     
For the symmetric case, the curve is       
\eqn\svthing{       
\eqalign{       
v &= {b\over {\lambda-1}} +  {b\over {\lambda+1}} \cr       
w &= {\mu b\over {\lambda-1}} -  {\mu b\over {\lambda+1}} \cr       
y &= A v^{2} \prod_{i=1}^{N_f} {\lambda^2 + 1 - \lambda c_i \over      
\lambda^2 -1} \left( {\lambda-1 \over \lambda+1} \right)^{2n_c-2-N_f} ~,\cr       
}}       
where we have defined       
\eqn\wehdef{       
\eqalign{       
c_i^2 &= {{4(b^2+m_i^2)}\over m_i^2} \cr       
b^{4n_c-4} &= (2\Lambda_{\CN=2})^{4n_c-4-2N_f}      
\prod_{i=1}^{N_f} m_i^2(1+c_i/2)^2\cr       
A^2 &= \Lambda_{\CN=2}^{4n_c-4-2N_f}\prod_{i=1}^{N_f} m_i^2 ~.\cr}   
}       
After a change of variables, these curves coincide with those found in     
\refs{\csst, \radu}       
where the related $SO$ and $Sp$ gauge theories were studied.       
%

%
%
%
%
        
\newsec{$\CN=1$ $SU(N_c)$ with an antisymmetric and $N_f=0,1$  
fundamental flavors}        
        
In the last section we have seen that the genus zero ansatz for         
the rotated curve associated with $\CN=1$ $SU(N_c)$ with a         
massive antisymmetric flavor and $N_f=0,1$ fails for generic        
values of the antisymmetric mass.         
\ifig\torus{This figure illustrates what happens when one tries to  
rotate the brane configuration for $SU(N_c)$ gauge theory with a massive flavor        
in the antisymmetric representation and no fundamental flavors.}        
{\epsfxsize=4truein        
\epsfysize=3.5truein        
\epsfbox{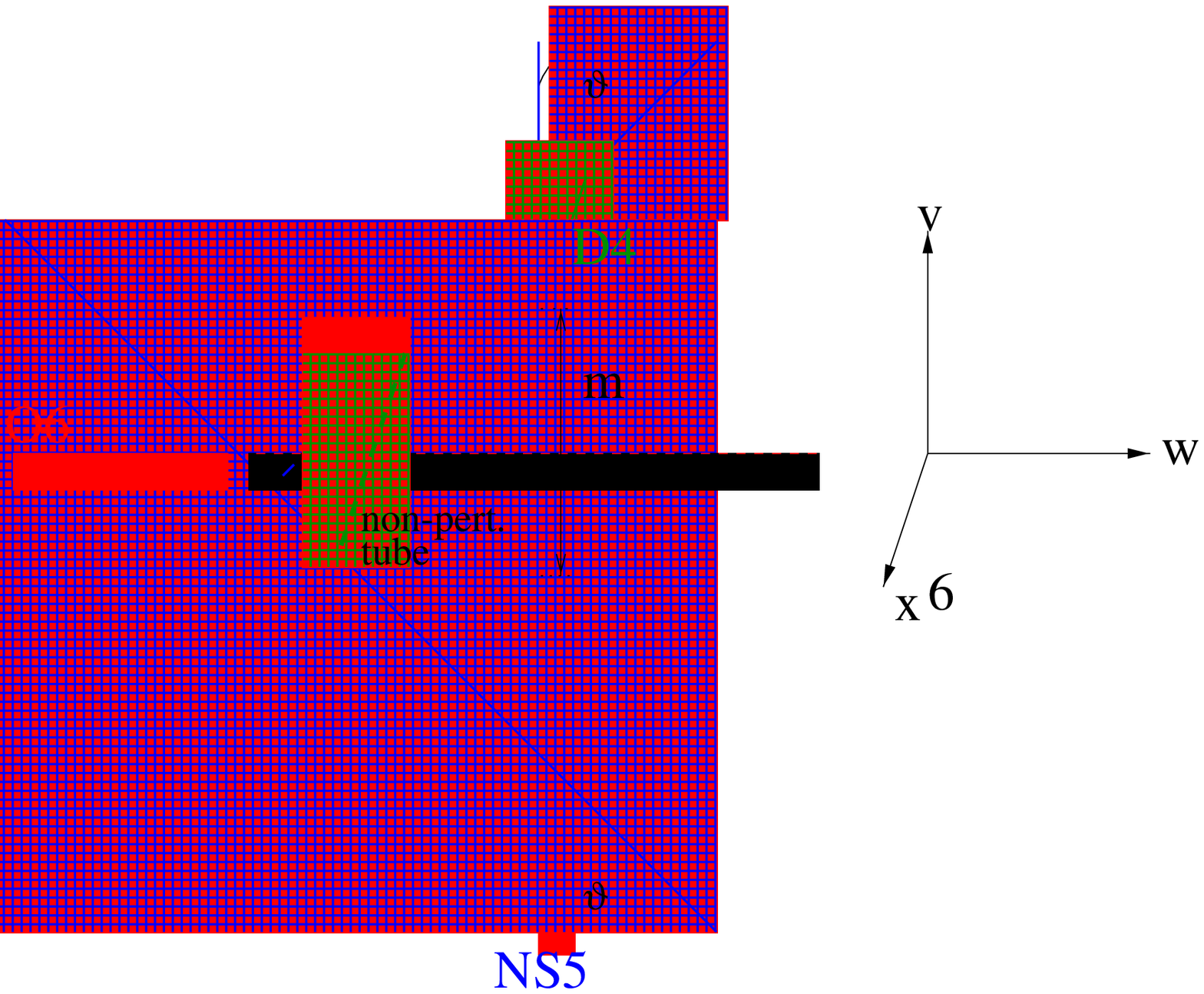}}        
The rotated brane configuration         
for $m\neq 0$ is shown in \torus. It is important to analyze it        
in a neighborhood of the point $(v=0,w={\mu m / 2})$.        
Around this point the configuration reduces to two fivebranes         
rotated symmetrically with respect to the orientifold sixplane.        
We recall now a crucial feature of the curves associated         
to $\CN=2$ theories with symplectic gauge group. For $N_f=0,1$  
they possess one more handle than expected from the number of         
fourbranes present \llli, \lalo. The additional        
handle originates from non-perturbative effects due to        
the orientifold plane and does not have a physical $U(1)$     
associated with it. This feature appears both        
when the symplectic projection is imposed by an orientifold        
fourplane of positive Ramond charge, or an orientifold         
sixplane of negative Ramond charge.        
Since this effect only depends on the orientifold plane,         
it must also be present for a configuration of two         
fivebranes and no fourbranes. We will assume that even        
when the fivebranes are rotated, a non-perturbative tube        
is generated connecting them.        
This suggests that the $\CN=1$ brane configuration of \torus\        
will define a genus one Riemann surface instead of a genus         
zero surface. The extra handle will come from the        
non-perturbative spike created by the orientifold sixplane        
of negative Ramond charge around $w= {\mu m / 2}$.        
        
We will analyze first the case $N_f=0$. Let us consider the        
asymptotic behavior that the rotated curve should have at        
$v\rightarrow \infty$ and $v=0$. It will be convenient to use         
the curves in the form derived in section 1, i.e. without         
rescaling the coordinate $y$ as we did in the previous section.         
The behavior as $v\rightarrow \infty$ should be         
\eqn\asym{        
\matrix{        
i) & y \rightarrow - v^{N_c}, \hfill& w \rightarrow  \mu v ,\hfill \cr     
ii) &  y \rightarrow (-1)^{N_c+1} v^{-2}, \hfill & w \rightarrow 0 ,\hfill \cr        
iii) & y \rightarrow (-1)^{N_c+1} v^{-N_c-4}, \hfill & w \sim -\mu v ~.    
\hfill\cr}        
}         
where $i)$, $ii)$ and $iii)$ correspond to the left,         
central and right fivebranes respectively. The behavior in a        
neighborhood of $v=0$ is fixed by the non-perturbative         
effects associated with the orientifold to be         
\eqn\asymzero{        
y \sim - v^{-2}~.        
}        
In the following we want to show that a genus one curve can        
satisfy \asym, \asymzero; conditions that a genus zero curve was        
unable to meet for generic $m \neq 0$.         
We will comment at the end about how to relate the elliptic modulus $\tau$        
of the genus one curve  with the gauge theory parameters.        
We know however that the specific values of $m$ derived from      
\specialzeroes\ should correspond to degenerations of the curve.        
        
A torus is defined by $C/L$, where $L$ is a lattice in the complex         
plane generated by vectors $(1, \tau)$. We denote by $z$        
the coordinate parameterizing a fundamental cell in $C/L$. We        
represent again the orientifold action by $z \rightarrow -z$.         
There are four fixed points under this involution:        
$0$, $1/2$, $\tau /2$, $(\tau + 1)/2$. Let us compactify the 
surface defined by the brane configuration by adding three points 
corresponding to the asymptotic regions $v \rightarrow \infty$
of each fivebrane. In analogy with the         
treatment of the genus zero curves, we want to construct now a         
holomorphic map from $C/L$ to the type \IIA\ ambient space
\eqn\torus{        
v=f_1(z) \;\;\; , \; w=f_2(z) \;\;\; , \;y=f_3(z)~.        
}        
The functions $f_i(z)$ will be meromorphic and doubly periodic.        
We begin by determining $f_1$. The function $f_1$ should have three    
simple poles representing the positions of the fivebranes. Since one       
fivebranes is its own mirror under the orientifold action, one of the      
poles should be at an invariant point under $z \rightarrow -z$. Let us     
denote it by $z_1$, where $z_1$ will be $0$, $1/2$, $\tau /2$ or         
$(\tau + 1)/2$; we call $z_2,z_3,z_4$ the other three invariant points.    
We set the other two poles at some points $z_0$, $-z_0$.         
A meromorphic function on a compact Riemann surface satisfies        
\eqn\orders{        
\sum_j m_j=0~,        
}        
where $j$ labels the zeroes and poles of the function and $m_j$ denotes    
its order at that points ($1$ if the function has a simple zero, $-1$ if   
it has a simple pole). Therefore $f_1$ will have in general three simple   
zeroes. We impose that they are at $z_2,z_3,z_4$. This determines the      
function $f_1$ up to a multiplicative constant.         
        
There is however an important additional constraint. Since the        
variable $v$ is odd under the orientifold involution, $f_1$ should  
be odd under $z \rightarrow -z$. In order to prove this we analyze         
${\tilde f}_1 = f_1(z)+f_1(-z)$. ${\tilde f}_1$ can have only   
simple poles at $\pm z_0$, however it still has zeroes at $z_2,z_3,z_4$.   
The only possibility compatible with \orders\ is that ${\tilde f}_1 =0$,   
and therefore the above defined function $f_1$ is odd.        
       
The asymptotics as $v \rightarrow \infty$ tells us that $f_2$ should       
have two simple poles at $\pm z_0$, and that $w\sim \mu v$ at $z_0$    
and $w\sim -\mu v$ at $-z_0$. This fixes $f_2$ up to shifts by a         
constant. Defining the function ${\tilde f}_2 = f_2(z)-f_2(-z)$        
and using similar arguments to those applied to ${\tilde f}_1$,        
one can see that ${\tilde f}_2=0$. Thus $f_2$ is an even function        
under the orientifold involution, as needed in order to map it to $w$.     
At this point we could explicitly construct $f_1$ and $f_2$ in terms of    
the Weierstrass functions. The function $f_3$ will prove however to be     
more involved. Therefore we keep at a more abstract level,        
determining each function in terms of the singularity structure we       
desire but without attempting to give their explicit form.        
       
On this line, we use both the asymptotic at $v \rightarrow \infty$  
and $v=0$ for determining $f_3$. Relations \asym, \asymzero\ imply    
that $f_3$ must have a pole of order $N_c$ at $z_0$, three poles        
of order $2$ at $z_2,z_3,z_4$, a zero of order $2$ at $z_1$        
and a zero of order $N_c+4$ at $-z_0$. These conditions fix         
$f_3$ up to a multiplicative constant.         
As it was the case with $f_1$ and $f_2$, $f_3$ should satisfy        
additional properties for correctly representing the coordinate        
$y$. First, invariance under the orientifold projection requires        
\eqn\fthree{        
f_3(-z) f_3(z) f_1^4 (z)= {\rm constant}~.        
}        
This is indeed satisfied because the left hand side is        
a function without poles, and the only function without        
poles in a compact Riemann surface is a constant.         
        
Next we have to check if $f_3$ fulfills \asym, \asymzero\ with        
the indicated proportionality coefficients. Multiplying $f_3$ and        
$f_1$ by appropriate constants, we can set the constant in         
\fthree\ to one and satisfy exactly conditions $i)$ and $iii)$.        
We have constructed the map \torus\ such that the asymptotics        
$ii)$ in \asym\ and \asymzero\ appear at the four invariant        
points of the torus under $z \rightarrow -z$. At these points         
\fthree\ reduces to        
\eqn\threepoint{        
f_3^2  = v^{-4}~.        
}        
Thus when $z \rightarrow z_i$, $i=1,..,4$, we get $y \rightarrow       
\epsilon v^{-2}$ with $\epsilon= \pm 1$. This is compatible with what    
we need, but still not a satisfactory answer. We need to obtain more    
information about the allowed distributions of $\epsilon$ values.        
        
In order to proceed further we define the function ${\tilde f}= f_3 +    
f_1^{-2}$. This function will have a different singular behavior         
depending on the value of $\epsilon$ at each point $z_i$.         
When $\epsilon=-1$ at one of the $z_i$, the singularity order of         
${\tilde f}$ at that point decreases with respect to that of $f_3$.        
One can see that the order $\tilde f$ at $z_i$ must be odd if        
$\epsilon_{z_i}=-1$ \foot{For proving this we need to define still       
one more auxiliary function: $f'_3$ such that $f_3=f'_3 f_1^{-2}$. At    
the invariant points $f'_3 = \epsilon + a z^n$. From \fthree\ we get     
$f'_3(z) f'_3(-z)=1$. This implies that $n$ has to be an odd integer.    
Since $f_1^{-2}$ is even under the orientifold involution, the order    
of ${\tilde f}$ at $z_i$ is odd if $\epsilon_{z_i}=-1$.}.        
The function ${\tilde f}$ has a pole of order $N_c$ at $z_0$        
and a zero of order $2$ at $-z_0$. It can also have        
zeroes at other points. Using again \fthree\ we derive the following    
equality       
\eqn\za{        
{f_3(z)+f_1^{-2}(z) \over f_3(-z)+f_1^{-2}(-z)}= f_1^2(z) f_3(z)~.    
}        
This implies that the additional zeroes of ${\tilde f}$ are at paired      
points $\pm z_a$, with each pair having the same order. Using all this     
information we can see that consistency fixes the value of        
$\epsilon_{z_1}$ in terms of the values of $\epsilon$ at $z_2,z_3,z_4$.    
According to \asymzero\ we are interested in $\epsilon=-1$ at         
$z_2,z_3,z_4$. Applying \orders\ to ${\tilde f}$ implies that         
$\epsilon_{z_1}=(-1)^{N_c+1}$, in agreement with condition $ii)$ in      
\asym.        
        
Finally, we will show that the distribution of $\epsilon$ values        
is correlated with the choice of point $z_0$. We will need to        
use an additional property satisfied by a meromorphic function on  
a torus        
\eqn\ma{        
\sum_j m_j z_j = 0 ~( {\rm mod}~ L)~.        
}        
Applying \ma\ to $f_3$ restricts the allowed values of $z_0$ to         
those satisfying $(2N_c+4)z_0 = 0 (\hbox{mod L})$. We see therefore      
that $z_0$ is not an additional modulus of our construction,        
but it is restricted to a discrete set of values. Applying \ma\         
to $\tilde f$ constrains further $z_0$. As a result, $\epsilon=-1$ at    
$z_2,z_3,z_4$ is only compatible with $(N_c+2) z_0 =0(\hbox{mod L})$     
for $N_c$ even and $(N_c+2) z_0 = z_1 (\hbox{mod L})$ for $N_c$ odd.     
Any other distribution of $\epsilon$        
values implies different values for $z_0$ \foot{With the         
exception of $\epsilon=1$ at $z_2,z_3,z_4$ and         
$\epsilon_{z_1}=(-1)^{N_c}$. These values just correspond to change $y   
\rightarrow -y$.}. Thus we can select the distribution of        
$\epsilon$ values compatible with the asymptotic behavior        
\asym, \asymzero\ by choosing appropriately $z_0$.        
        
We would like to comment on how to determine the elliptic modulus of       
the $\CN=1$ curve. We could expand $f_1$ and $f_3$ around $z_1,\pm z_0$ and         
substitute in the $\CN=2$ curve, as we did for the genus zero curves.    
In this way we would obtain relations between the coefficients of the      
subleading terms in $f_3$ and $f_1$ and the mass of the antisymmetric      
flavor. Since $f_3,f_1$ are fixed, these subleading coefficients         
are functions of the elliptic modulus $\tau$ and $z_0$. Therefore we       
will have relations between $\tau$, $z_0$ and $m$, as we wanted. To be able        
to find a consistent set of relations for the subleading terms would be    
a further test of our genus one curve. This could restrict further the     
allowed values of $z_0$. The set of allowed $z_0$ should provide the    
$\CN=1$ vacua of our theory.       
        
The construction of genus one curve associated with the $\CN=1$ $SU(N_c)$  
theory with a massive antisymmetric flavor and $N_f=1$ can be done    
in a completely analogous way \foot{We are referring again to a flavor  
induced by a pair of sixbranes placed over the orientifold sixplane,  
at $v=0$.}.       
The functions $v=f_1(z)$ and $w=f_2(z)$ will have the same singularity        
structure as before. The function $y=f_3(z)$ should have now a pole of   
order $N_c$ at $z_0$, a zero of order $N_c+2$ at $-z_0$, a simple       
zero at $z_1$ and simple poles at $z_2,z_3,z_4$. This function satisfies   
a version of \fthree\       
\eqn\fthreeone{       
f_3(z) f_3(-z) f_1^2 (z) = {\rm constant}.       
}       
From this one deduces $f_3 \rightarrow \epsilon v^{-1}$ when       
$z \rightarrow z_i$ with $\epsilon =\pm 1$. The $\CN=2$ curve for    
$SU(N_c)$ with an antisymmetric flavor and $N_f=1$ \antisymF\       
reduces at $v=0$ to $(y+v^{-1})^2 (y-v^{-1})=0$. We       
therefore set $\epsilon=-1$ at  $z_2,z_3$, and        
$\epsilon=1$ at $z_4$. It is convenient to define now ${\tilde f}= f_3        
+ f_1^{-1}$. Using the same considerations as previously one can see    
that the above values of $\epsilon$ at $z_2,z_3,z_4$ are only        
compatible with $\epsilon=(-1)^{N_c+1}$ at $z_1$. In agreement with this,        
the $\CN=2$ curve at the pole associated with the central fivebrane      
behaves as $y \rightarrow (-1)^{N_c+1} v^{-1}$.        
Finally, we can isolate as before the desired distribution of       
$\epsilon$ values by an appropriate choice of the point $z_0$.

\newsec{Chiral Theory}        
        
In this section we discuss the curve for the chiral theory with $SU(N_c)$       
gauge group, $2N_f+8$ chiral multiplets in the fundamental representation,     
$2N_f$ in the antifundamental, one in the antisymmetric and one in the     
conjugate symmetric.     
\ifig\figiv{Brane configuration for the chiral theory.}        
{
\epsfxsize=3.5truein\epsfysize=3truein        
\epsfbox{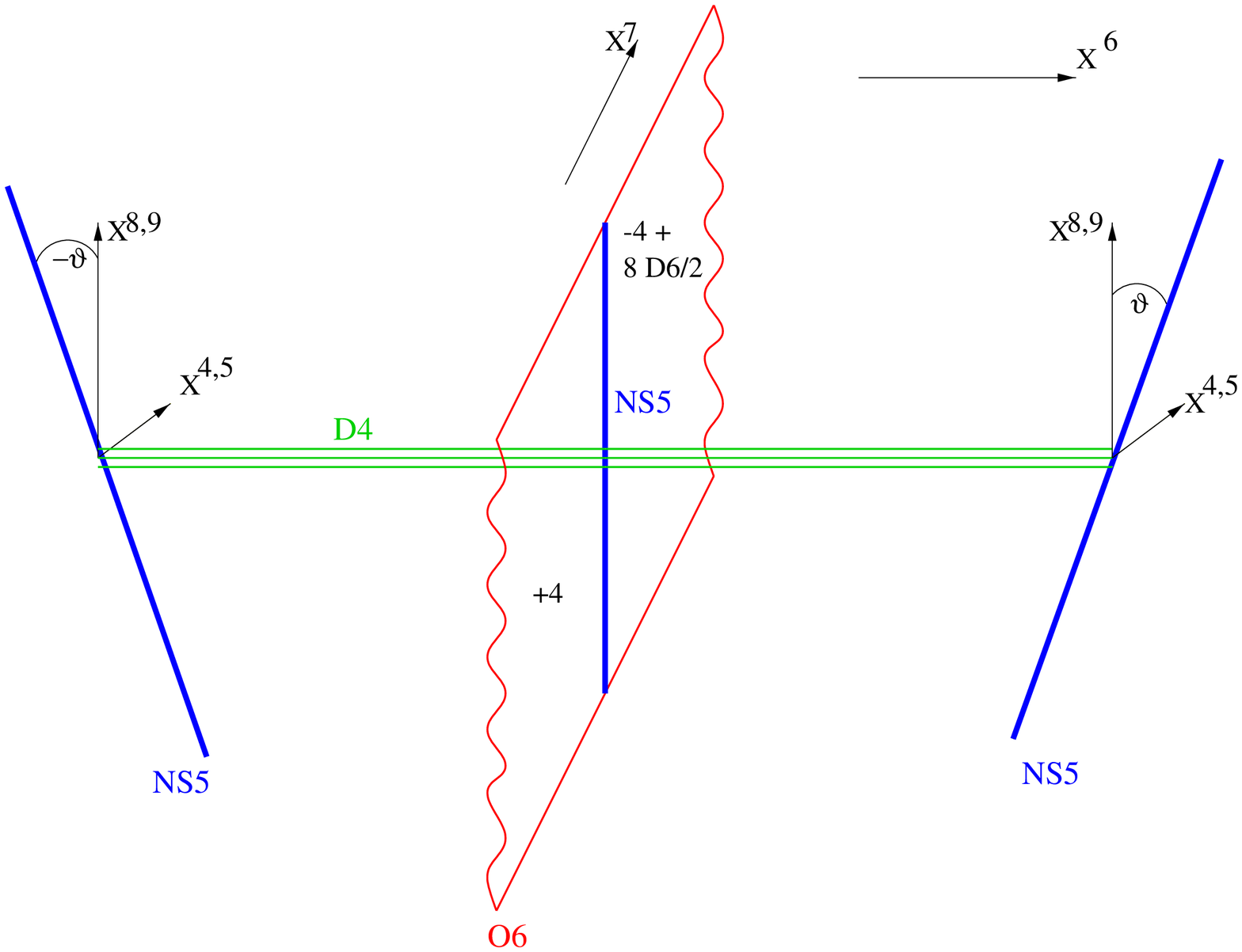}}    
    
The brane construction that gives     
rise to this theory has previously been studied in \lllii.     
It is again a    
configuration containing three fivebranes in the background of an    
orientifold sixplane. There are only two possible orientations for the     
middle fivebrane, which is its own mirror under the orientifold action.    
It can extend in the $v$-plane, as in the previous sections, or it can  
extend in the $w$-plane. This second possibility gives raise to the    
chiral theory. The orientifold sixplane is then divided in two by the    
middle fivebrane. At this point, the sign of the orientifold projection    
must change \evans. In order to compensate the change in Ramond charge  
of the orientifold, we have to introduce  an additional set of 8 half  
D-sixbranes parallel to the orientifold plane, ending on the central     
fivebrane (see \figiv). These give rise to the additional eight     
fundamentals \haza. Strings joining the sets of fourbranes to the left  
and right of the middle fivebrane induce now a chiral multiplet in the  
antisymmetric representation and an one in the conjugate symmetric    
representation.     
    
We include also the presence of $2N_f$ sixbranes parallel to the     
orientifold sixplane. If the sixbranes are placed on top of the orientifold     
they get also cut by the middle fivebrane. In this special situation the   
matter content they induce on the gauge theory living on the fourbranes    
is doubled \lllii. The flavor symmetry group is enhanced from     
$SU(N_f)_L \times SO(8)_L\times SU(N_f)_R$ to     
$SO(2N_f+8)_L \times Sp(2N_f)_R$.    
As before we can rotate the two outer fivebranes in a symmetric way    
with respect to the middle one by an angle $\theta$.     
The tree-level superpotential for this brane configuration is     
\eqn\chisuper{     
W = Q {\tilde X}_s  Q   + \tilde Q X_a \tilde Q + X_a X {\tilde X}_s +   
\mu X^2 ~,    
}     
where $\mu=\tan \theta$, $X$ is the adjoint multiplet, ${\tilde X}_s$  
the symmetric, $X_a$ the antisymmetric, $\tilde Q$ the antifundamentals  
and $Q$ the fundamentals.         
    
The chiral brane configuration is such that the fourbranes to the left  
and right of the middle fivebrane can always be reconnected. Thus we    
expect that the associated curve factorizes into two pieces for any value    
of the parameters. The first piece will be a $\CP^1$ representing    
the middle fivebrane as before. The second piece should describe a    
configuration with two fivebranes in a background space with a    
uniform $+4$ Ramond charge along the directions $(0123678)$. Thus    
for $\mu=\infty$ this curve will coincide with that describing an     
$\CN=2$ $SO(N_c)$ theory with $N_f$ massless flavors or equivalently,    
if $N_c$ is even, an $Sp(N_c)$ theory with $N_f+4$ massless flavors.     
When $\mu$ is finite, the curve for the chiral theory will coincide with   
the rotated curve for the $SO$ or $Sp$ theories with tensor matter mentioned    
above.    
    
This chiral theory contains baryon operators $\tilde B_n=\tilde        
X^n_s \tilde Q^{N_c-n} \tilde Q^{N_c-n}$ and $B_n = X^n_a Q^{N_c-2n}$.   
By moving the central fivebrane in the positive $x^7$ direction we move    
onto the $B_{n}$ baryonic branch and the gauge group breaks to $Sp(n)$ with      
$2N_f+8-2(N_c-n)$ massless chiral fundamentals     
(with $SO(2N_f+8-2(N_c-n))$ flavor symmetry) \lllii. Likewise, when we move     
the     
fivebrane in the negative $x^7$ direction, we move onto the $\tilde B_n$   
baryonic branch and the gauge group breaks to $SO(n)$ with $2N_f-2(N_c-n)$     
massless fundamentals, with $Sp(2N_f-2(N_c-n))$ flavor symmetry.     
The curve we have presented describes the chiral     
theory at the $B_{N_c}$, $\tilde   B_{N_c}$ baryonic branch         
root. The other $B_n$, $\tilde B_n$         
baryonic branches will be obtained by rotating        
the $\CN=2$ curve at the origin of the lower baryonic branches.        
    
As in the $\CN=2$ case there is a $U(1)_X$, that now acts on $X_a$, 
$\tilde X_s$, $Q$ and $\tilde Q$. Its Goldstone mode 
together with the $x^7$ position of the detached fivebrane gives rise to
one complex modulus. This     
complex modulus appears to be frozen from the M-theory point of view since   
the fivebrane in question appears to be of infinite size.  It is possible
when   
one flows to the field theory limit, there are corrections to the Kahler   
potential, and the metric is no longer degenerate in this direction.


        
\newsec{Higgsing for $\CN=2$}       
     
The Higgs branch appears when the M5-brane intersects a singular point     
in the multi-Taub NUT space. This point can be resolved into a number    
of rational curves which are then free to move off in the $7,8,9$       
directions \witM, \hoo. The Higgs moduli correspond to these parameters    
together with their superpartner, arising       
from integrating the chiral two-form of the worldvolume theory of the    
fivebrane over the rational curve.      
       
\subsec{Resolution of $D_n$ Singularity}       
       
We begin by reviewing the minimal resolution of a $D_n$ singularity  
\eqn\dneqn{       
a^2 +b^2 z = z^{n-1}~.       
}       
The resolved surface is covered by $n$ open sets $U_1, \cdots U_n$       
with coordinates $(s_1, t_1, z_1)=(a,b,z/a)$,        
$(s_2=b, t_2=a/z, z_2),\cdots, (s_n,t_n,z_n)$. These are glued       
together via the transition relations       
\eqn\transrel{       
\eqalign{       
(s_j,t_j, z_j) &= (s_{j+1} t_{j+1} z_{j+1}, s_{j+1}, t^{-1}_{j+1} )      
\quad j=1,\cdots,n-4 \cr        
(s_{n-3},t_{n-3}, z_{n-3} ) &= (s_{n-2} t^2_{n-2} z_{n-2}, s_{n-2}    
t_{n-2}, t^{-1}_{n-2}), \cr        
(s_{n-2},t_{n-2}, z_{n-2} ) &= ( t_{n-1} z_{n-1}, s_{n-1},        
t^{-1}_{n-1} ) = (t_n^{-1}, s_n t_n, z_n) ~,\cr}        
}        
and the projection to the $a,b,z$ space is        
\eqn\projec{        
\matrix{        
a &= s_{2j-1}^j z_{2j-1}^{j-1} \hfill  &= s^j_{2j} t_{2j} z_{2j}^j \hfill\cr        
b &= s_{2j-1}^{j-1} t_{2j-1} z_{2j-1}^{j-1}\hfill &= s_{2j}^j z_{2j}^{j-1}        
\hfill\cr        
z &= s_{2j-1} z_{2j-1}\hfill &= s_{2j} z_{2j}~,\hfill \cr}        
}        
for $U_1,\cdots, U_{n-3}, U_{n-1}$. For $U_{n-2}$ and $U_n$ we have        
\eqn\unprojec{        
\matrix{        
a &= z^{n/2-2} s_{n-2} t_{n-2} \hfill &= s_n z^{n/2-2}, \hfill & n: ~{\rm even}        
\cr        
b &=  z^{[n/2] -1} t_{n-2} \hfill &= s_n t_n z^{[n/2]-1},        
\hfill        
& n: ~{\rm        
odd} \cr        
z&= s_{n-2} t_{n-2} z_{n-2}\hfill &= s_n z_n ~.\hfill& \cr}        
}        
        
The inverse image of the singular point consists of $n$ rational        
curves $C_i$. For $i=1,\cdots , n-2$ these are the $z_i$ axis in $U_i$   
the $t_{i+1}$ axis in $U_{i+1}$. $C_{n-1}$ and $C_n$ are the        
curves $t_{n-2} = z_{n-2} \mp 1 =0$ in $U_{n-2}$. The $D_n$ singularity in        
the $i$-th patch takes the form $s_i + t_i^2 z_i = s_i^{n-i-1} z_i^{n-i}$,        
$s_{n-2}+t_{n-2} z_{n-2} = s_{n-2} z_{n-2}^2$ in $U_{n-2}$ and        
$1 + s_n t_n^2 z_n = z_n^2$ in $U_n$. Close to the exceptional divisor   
$C_i$ we have $s_i,t_i \to 0$ and from the form of the $D_n$ singularity   
in this patch $s_i \sim t_i^2$ while close to $C_{i+1}$ $s_i,z_i \to 0$ with     
$s_i \sim z_i$.       
It is important to take the latter into consideration      
when counting the multiplicities       
of the $C_i$'s on the Higgs branches.        
      
\subsec{$Sp(2N_c)$ Gauge Group}       
        
We consider the gauge group $Sp(2N_c)$ with $N_f$ fundamental        
flavors. The singularity is of type $D_{N_f}$.  The curve is given by      
$b=P(z)$ where $P(z) = z^{N_c} + u_2 z^{N_c-1} + \cdots$.  As long as        
$r \neq N_f/2$ (for the case $N_f$ even),         
we can assume that at the $r$-th        
Higgs branch root         
\eqn\spbranchrooti{ P(z) = z^r \tilde P(z)\,.}         
Away from        
$(a,b,z)=(0,0,0)$ we can rescale $(a,b) \rightarrow (z^r a, z^r b)$ and        
then we recover the curve for $Sp(2(N_c-r))$ + $N_f-2r$ flavors, with      
$r \leq \hbox{min}(N_c,\left[ {N_f/ 2}\right])$.         
Close to the singularity the curve takes the form $b = z^r$. This        
equation has to be analyzed now in each of the coordinate patches defined        
above.        
On the rational curves         
$C_1, \cdots, C_{N_f}$ one finds         
$1,2,\cdots, 2r-1, 2r, \cdots, 2r,r,r$ solutions        
respectively. In addition, one finds an infinite component that        
intersects the curve $C_{2r}$. From this it follows that the quaternionic dimension of        
the Higgs branch is given by $2r N_f - r (2r+1)$ in agreement with field   
theory. This has also already been discussed in a slightly different  
context in \hov.         
        
We have to be more careful when $N_f$ is even and $r={N_f / 2}$.  In  
this case we are left with the unbroken gauge group $Sp(2N_c -N_f)$ and no        
flavors. This curve does not live in a $D_n$ space but in the        
Atiyah-Hitchin manifold         
\eqn\ath{ a^2 + b^2 z = 2b\,.}         
To go from        
$D_{N_f}$ to Atiyah-Hitchin we have to redefine        
$(a,b) \rightarrow (z^{N_f/2}, z^{N_f/2-1}(bz-1) )$.         
From this it follows that the polynomial        
$P(z)$ at the $N_f/2$-th Higgs branch root has to factorize as        
\eqn\spbranchrootii{ P(z) = z^{N_f/2-1}(\tilde P(z) z-1) \,.}         
We see that        
the location of this Higgs branch is shifted. It is rather nice to see how        
this shift appears through geometrical considerations by going from a      
$D_{N_f}$ singularity to Atiyah-Hitchin. This constitutes another fine  
example for the interplay between geometry and gauge theory. In a        
different setup with an orientifold four-plane this shift has also found   
in \dhoo.  In the odd coordinate patches the curve takes the form        
\eqn\curveoddpatches{ s_{2j-1}^{j-1} z_{2j-1}^{j-1} \left( t_{2j-1} -      
s_{2j-1}^{N_f/2-j} z_{2j-1}^{N_f/2-j} (s_{2j-1} z_{2j-1}-1) \right)=0\,,}        
for $j\leq N_f/2-2$ and in the $N_f-2$-th patch it is         
\eqn\curvelastpatch{        
s_{N_f-2}^{N_f/2-1} t_{N_f-2}^{N_f/2-1} \left( 1 - z_{N_f-2} (s_{N_f-2}    
t_{N_f-2} z_{N_f-2} -1)\right) =0 \,.}         
The exceptional divisors appear        
with multiplicity $i$ for $i=1\cdots,N_f-2$ with multiplicity        
${N_f/ 2}-1$ for $i=N_f-1$ and ${N_f/ 2}$ for $i=N_f$. All together    
this gives the correct dimension of the Higgs branch         
${N_f} (N_f-1)/2$.

\subsec{$SO(2N_c)$ Gauge Group}        
        
The gauge group $SO(2N_c)$ is obtained with a orientifold sixplane  
with $+4$ units of Ramond charge. The singularity is now of type        
$D_{N_f+4}$. This can be shown as in \torused\ by starting        
with the orientifold description as in \xyspace\ and then introducing      
invariant variables $a=v(x-y)$, $b=x+y$, $z=v^2$.        
We obtain then the $D_{N_f+4}$ singularity with the additional restriction        
that we can only resolve down to $D_4$, which by itself represents the     
orientifold. The curve is given by $b = P(z)$.        
First consider the case where $SO(2N_c)$ with $N_f$        
flavors is broken down        
to $SO(2(N_c-r))$ with $N_f-2r$ flavors.        
Near the singularity the curve is described by        
$b-z^r=0$. We need to consider blow ups which leave a $D_4$        
singularity intact. This corresponds to blowing up only the first        
$N_f$ rational curves that appear in the resolution of a $D_{N_f+4}$    
singularity. The multiplicities are identical to those found above,        
namely $1,2,\cdots, 2r-1$ for $C_1, \cdots ,C_{2r-1}$ and         
$2r$ for $C_{2r}, \cdots C_{N_f}$. Summing        
the total number of rational curves gives $2r N_f - 2 r^2 +r$ for the      
quaternionic dimension of the Higgs branch, in agreement with field        
theory.        
        
It is also possible to consider Higgsing $SO(2N_c)$ with $N_f$        
flavors down        
to $SO(2N_c-2r-1)$ with $N_f-2r-1$ flavors. This can be understood as      
a sub-branch of the previous case, except when $N_f$ is odd and        
$r=[N_f/2]$.        
Now we assume that $P(z)=z^{r+1} \tilde P(z)$. To obtain the curve away from        
the singularity we rescale $b \rightarrow z^r a$ and $a \rightarrow z^{r+1} b$.        
These rescalings follow from demanding that we have a $D_4$         
singularity $ z b^2 + a^2 = z^3$ remaining. Note that the curve  
is described by         
$a= z \tilde P(z)$. Relative to the previous case, the roles played by the         
coordinates $a$ and $b$ are interchanged.        
Curves for $SO(2N_c)$ are        
described by $b=P(z)$ in a $D_{N_f+4}$ space \dneqn\ whereas curves for        
$SO(2N_c+1)$ are described by $a= z P(z)$ with $P$ being a polynomial of        
order $N_c$. We will explain this immediately in        
the next subsection.        
The curve near the singularity        
takes the form $a - z^{1+r}=0$. Analyzing this on the patches        
$U_1,\cdots, U_{N_f}$ one finds the solutions on the curves         
$C_1,\cdots, C_{N_f}$ with multiplicities        
$2,3,\cdots,2r+1,2r+2,\cdots 2r+2$. On each of these $C_i$ one of       
these solutions corresponds to an additional infinite        
D-fourbrane along the $z$        
axis. To Higgs to $SO(2N_c-2r+1)$ we want to keep this D-fourbrane       
intact. The quaternionic dimension of the moduli space is then       
$\sum_{i=1}^{2r+1} i  + (2r+1)(N_f-2r-1) = (2r+1)(N_f-r)$.       
       
\subsec{$SO(2N_c+1)$ Gauge Group}        
        
Our first task is now to understand the particular form for the curves.    
We start with the description in terms of $x$, $y$ and $v$,        
\eqn\oddso{ xy = v^{4+2N_f}\qquad y^2 + y v P(v^2) - v^{4+2N_f} =0\,.}        
This can be written as $ y - x = v P(v^2)$. In order to write this in    
terms of invariant variables we have to multiply with an overall factor    
of $v$. This means that we add an additional infinite fourbrane         
at $v=0$. Now the curve can be written in a $D_{N_f+4}$ space as $a = z P(z)$.         
This explains the form that we obtained by Higgsing previously.        
Assuming $P(z) = z^r \tilde P(z)$ we get the curve away from the singularity       
by taking $(a,b) \rightarrow (z^{r+1} b,z^r a)$ and then we recover     
the curve for $SO(2N_c-2r)$ gauge theory with $N_f-2r-1$ flavors.    
    
We can now        
go on and count the multiplicities of the exceptional divisors in the blown up       
space.        
Close to the singularity the curve takes the form $a=z^{r+1}$.        
In the $i$-th patch we find the exceptional divisor with multiplicity      
$i+1$ for $i=1, \cdots, 2r$ and with multiplicity $2r+2$ for $i=2r+1, \cdots,        
N_f$. Now we have to remember that we had to add an additional fourbrane   
to write the curve in invariant variables. This fourbrane does not contribute        
to the Higgs branch of the gauge theory. However, we still should expect to        
see it in all the patches. Thus in order to compute the dimension we should        
subtract one from the multiplicities in each patch. We get then $N_f (2r+1)-        
r(2r+1)$ which indeed is the correct quaternionic dimension of the Higgs   
branch.     
    
The cases with unbroken gauge group $SO(2(N_c-r)+1)$ and $N_f-2r$ flavors can be       
understood as sub-branches of the previous case except for $N_f$ even and       
$r=N_f/2$, when we break to the theory with no flavors.     
We assume $P(z) = z^{N_f/ 2} \tilde P(z)$. To get the curve        
away from       
the singularity we have to rescale $(a,b) \rightarrow (z^{N_f/ 2} a,    
z^{N_f/ 2} b)$. Close to the singularity the curve is again $a=       
z^{{N_f/ 2}+1}$. Counting the multiplicities in the same way as before     
we get for the dimension of the Higgs branch ${N_f}(N_f+1)/2$.

\newsec{Higgsing for $\CN=1$}         
         
The Higgs branch for $\CN=2$ is typically only a subspace of the full    
Higgs branch of the $\CN=1$ $SO/Sp$ theory with tensors.          
The most interesting case to         
consider is when we send $\mu \to \infty$. The outer NS-fivebranes are     
parallel to the orientifold sixplane, and new moduli open up         
corresponding to fourbranes moving along the outer fivebranes.         
Additional moduli also arise from an infinite component of the $\CN=1$   
curve degenerating as $\mu\to \infty$ into another infinite curve plus     
rational curves.         
         
\subsec{$Sp(2N_c)$ Gauge Group}         
         
For $\mu \to \infty$ we have an $Sp(2N_c)$ gauge theory with a         
massless antisymmetric tensor and $N_f$ flavors of fundamental. The limit         
is taken by rescaling $\tilde y = \mu^{2N_c} y$, $\tilde x = \mu^{2N_c} x$. The scale         
$\Lambda_{\CN=1}$ is held fixed in this limit, and is related to the     
$\CN=2$ scale by         
\eqn\lammer{         
\mu \Lambda_{\CN=2} = (\Lambda_{\CN=1}^{4N_c +8 -2N_f}          
\mu^{ -2N_f+4} )^{1\over{4N_c+4-2N_f}} ~.       
}         
Note that as we send $\mu\to \infty$ we not only integrate out the massive       
adjoint chiral multiplet of the $\CN=2$ theory, but we integrate in a    
light antisymmetric tensor with mass $1/\mu$. This spoils the naive dimension        
counting in \lammer. There are        
three cases to consider, $N_f<2$, $N_f=2$ and         
$N_f>2$. In the first case, $\mu \Lambda_{\CN=2}$ blows up as          
$\mu \to \infty$ and the curve becomes infinitely stretched in the         
$x^6$ direction. This is the manifestation in M-theory of the fact         
that the field theory has a runaway vacuum state \spanti.          
         
For $N_f>2$, $\mu \Lambda_{\CN=2}$ vanishes as $\mu \to \infty$  
and the $\CN=1$ curve        
splits into an infinite component       
\eqn\infcomp{\eqalign{         
C &: \quad v=0, \quad \tilde x=0, \quad \tilde y = w^{2N_c}~,\cr  }        
}        
and its $\IZ_2$ symmetric partner, where $\tilde x$ and $\tilde y$ are interchanged.       
In addition, there are       
a number of rational curves. These arise from the         
same scaling limit as $C$, namely $\mu \to \infty $, with           
$w^2\sim (\mu v)^2$ held finite.         
       
We wish to analyze the equations for the curve in the resolved $D_{N_f}$   
space as $\mu \to \infty$. Up to trivial rescalings, these equations reduce to       
\eqn\invvari{       
\eqalign{       
a &= v(\tilde y-\tilde x) = w^{2N_c+1}/\mu \cr       
b & = (\tilde x+\tilde y) = w^{2N_c} \cr       
z & = -(w/\mu)^2 ~.\cr}       
}       
On the $U_{2j}$ patch       
this yields         
\eqn\spnone{         
\eqalign{         
s_{2j} &= \mu^{2j-2} w^{2N_c-2j+2}  \cr         
z_{2j} &= \mu^{-2j} w^{2j-2N_c} ~.\cr}         
}         
For $j>N_c$ these equations can never be satisfied for finite $w$ and      
$s_{2j}$ so the curve does not have a solution in this patch. For          
$j\leq N_c$, we find solutions with $2(N_c-j)$ multiplicity on the         
curve $C_{2j}$. Likewise $2(N_c-j)+1$ solutions are found on         
$C_{2j-1}$.         
The total number of additional rational curves that appear is         
$\sum_{i=1}^{2N_c} (2N_c -j) =2 N_c^2 -N_c$. This gives rise to        
$4N_c^2-2N_c$ complex moduli.         
         
In addition, there are an extra $2N_c$ complex moduli arising from the     
blowup of the infinite component $C$ to $\tilde y = w^{2N_c} + a_1       
w^{2N_c-1} + \cdots + a_{2N_c}$. The total number of complex moduli is     
then that of the $\CN=2$ theory, plus the contributions from these two   
other sources. This totals $4 N_f N_c - 2 N_c$, in agreement with the      
field theory result \csst.          
         
For the special case $N_f=2$ a similar         
calculation leads to the same result for the dimension of the Higgs    
branch. In this case, $\mu \Lambda_{\CN=2}$ is held          
finite and the form of the curve in the $\mu\to \infty$ limit differs      
from \infcomp.

\subsec{$SO(2N_c)$ Gauge Group}         
         
The $SO(2N_c)$ case may be treated in a similar way.          
Now the scale         
$\Lambda_{\CN=1}$  is related to the         
$\CN=2$ scale by         
\eqn\lammerso{         
\mu \Lambda_{\CN=2} = (\Lambda_{\CN=1}^{4N_c -8 -2N_f}          
\mu^{ -2N_f-4} )^{1\over{4N_c-4-2N_f}} ~,        
}         
        
In the $\mu \to \infty$ limit $\mu \Lambda_{\CN=2}$         
vanishes and the $\CN=1$ curve       
splits into an infinite component       
\eqn\infcompso{\eqalign{         
C &: \quad v=0, \quad \tilde x =0, \quad \tilde y = w^{2N_c}~,\cr }         
}         
an its $\IZ_2$ image with $\tilde x$ and $\tilde y$ interchanged. There is also        
a number of rational curves, which arise from the         
same scaling limit as $C$; $\mu \to \infty $, with           
$w^2\sim (\mu v)^2$ held finite.          
         
The counting of the rational curves proceeds as in the previous subsection,        
with the only difference being that now the curve sits in the resolved $D_{N_f+4}$ space.       
$4 N_c^2 -2N_c$ extra complex moduli come          
from these curves, and $2N_c$ come from the blow up of the infinite    
component $C$. The total number of complex moduli for the Higgs         
branch is therefore $2(2N_c N_f - 2 N_c^2 +N_c)$ from the $\CN=2$ branch         
plus these additional contributions, which totals $4 N_c N_f +         
2N_c$. This agrees with the field theory result.

\subsec{$SO(2N_c+1)$ Gauge Group}

The $\mu \to \infty$ limit of the curve takes the form         
\eqn\soinflim{         
\eqalign{         
C &: \quad v=0, \quad \tilde x=0, \quad \tilde y = w^{2N_c+1}~,\cr}         
}         
plus its $\IZ_2$ image,       
plus rational curves, which arise from the same scaling limit as         
$C$, $w^2 \sim (\mu v)^2$.         
          
To write the curve in $a,b,z$ coordinates of $D_{N_f+4}$ space,        
we add an extra infinite D-fourbrane by multiplying through by        
an extra power of $v$ as        
explained in the section 6.4.       
The curve as $\mu\to \infty$ is taken to be       
\eqn\sooddcrp{       
\eqalign{       
a &= v(\tilde y-\tilde x) = w^{2N_c+2} \cr       
b & = (\tilde x+\tilde y) = \mu w^{2N_c+1} \cr       
z & = -(w/\mu)^2 ~.\cr}       
}       
Analyzing these equations       
in the $U_{2j}$ patch, one finds $2j$ solutions on the $C_{2j-1}$  
curve for $2j < 2N_c+2$. On $C_{2j}$ one finds $2j+1$ solutions for    
$2j< 2N_c+1$. One of these solutions on each $C_i$ corresponds to the      
additional infinite D-fourbrane along the $z$-axis, which appeared in      
the $\CN=2$ analysis. The number of additional rational curves is    
therefore $\sum_{i=1}^{2N_c} i = N_c (2N_c+1)$. There are also         
$2N_c+1$ complex moduli arising from the blowup of the infinite         
component $C$. The total number of additional complex moduli is         
$(2N_c+1)^2$, which when combined with the moduli of the $\CN=2$         
analysis, gives $2(2N_c+1) N_f + (2N_c+1)$. This agrees with the field     
theory analysis.

\vskip3cm        
\centerline{\bf Acknowledgments}        
K.L and E.L. would like to thank N. Nikbakht-Tehrani and M. Kreuzer for    
discussions. D.L. thanks the IAS and ITP, Santa Barbara for hospitality.   
The research of K.L. was supported by the FWF Project Nr. P10268-PHY,  
that of E.L. by the FWF, Lise Meitner Fellowship M456-TPH, that of     
D.L. by DOE grant DE-FE0291ER40688-Task A.

\listrefs        
\end